\documentclass[american,english]{IEEEtran}
\usepackage[T1]{fontenc}
\usepackage[latin9]{luainputenc}
\usepackage{color}
\usepackage{array}
\usepackage{float}
\usepackage{textcomp}
\usepackage{multirow}
\usepackage{amstext}
\usepackage{graphicx}

\makeatletter

\providecommand{\tabularnewline}{\\}

\newenvironment{lyxlist}[1]
	{\begin{list}{}
		{\settowidth{\labelwidth}{#1}
		 \setlength{\leftmargin}{\labelwidth}
		 \addtolength{\leftmargin}{\labelsep}
		 }}
	{\end{list}}

\usepackage{xtab,booktabs}
\usepackage[linesnumbered,ruled,lined,boxed]{algorithm2e}

\makeatother

\usepackage{babel}
\begin{document}
\title{Collective Intelligence: Decentralized Learning for Android Malware Detection in IoT with Blockchain}
\author{Rajesh Kumar, WenYong Wang,  Jay Kumar,  Zakria, Ting Yang \&  Waqar Ali  
\thanks{Manuscript received March 24, 2021; revised April 24, 2021; accepted
	April 24, 2021. This work was supported in part by the National Natural
	Science Foundation of China under Grant U2033212 
	The associate editor coordinating the review of
	this article and approving it for publication was xyz.
	(Corresponding author: Wenyong Wang.)}
	\IEEEcompsocitemizethanks{\IEEEcompsocthanksitem R.Kumar and J Kumar are with Yangtze Delta Region Institute (Huzhou), University of Electronic Science and Technology of China, Huzhou 313001, China, \protect\\
	E-mail: rajakumarlohano@gmail.com,  jay\_tharwani1992@yahoo.com
	\IEEEcompsocthanksitem W. Wang is with International Institute of Next Generation Internet, Macau University of Science and Technology, Taipa 999078, Macau SAR. E- mail: wywang@must.edu.mo
	\IEEEcompsocthanksitem T. Yang, W. Ali  and Zakria are University of Electronic Science and Technology of China, Chengdu, 611731, China	E-mail: abdkhan@hotmail.com,  zakria.uestc@hotmail.com
	
}}

\maketitle
\begin{abstract}
The widespread significance of Android IoT devices is due to its flexibility and hardware support features which revolutionized the digital world by introducing exciting applications almost in all walks of daily life, such as healthcare, smart cities, smart environments, safety, remote sensing, and many more. Such versatile applicability gives incentive for more malware attacks. In this paper, we propose a framework which continuously aggregates multiple user trained models on non-overlapping data into single model. Specifically for malware detection task, (i) we propose a novel user (local) neural network (LNN) which trains on local distribution and (ii) then to assure the model authenticity and quality, we propose a novel smart contract which enable aggregation process over blokchain platform. The LNN model analyzes various static and dynamic features of both malware and benign whereas the smart contract verifies the malicious applications both for uploading and downloading processes in the network using stored aggregated features of local models. In this way, the proposed model not only improves malware detection accuracy using decentralized model network but also model efficacy with blockchain. We evaluate our approach with three state-of-the-art models and performed deep analyses of extracted features of the relative model. 
\end{abstract}


\section{Introduction}

\begin{figure}
\centering{}\includegraphics[scale=0.06]{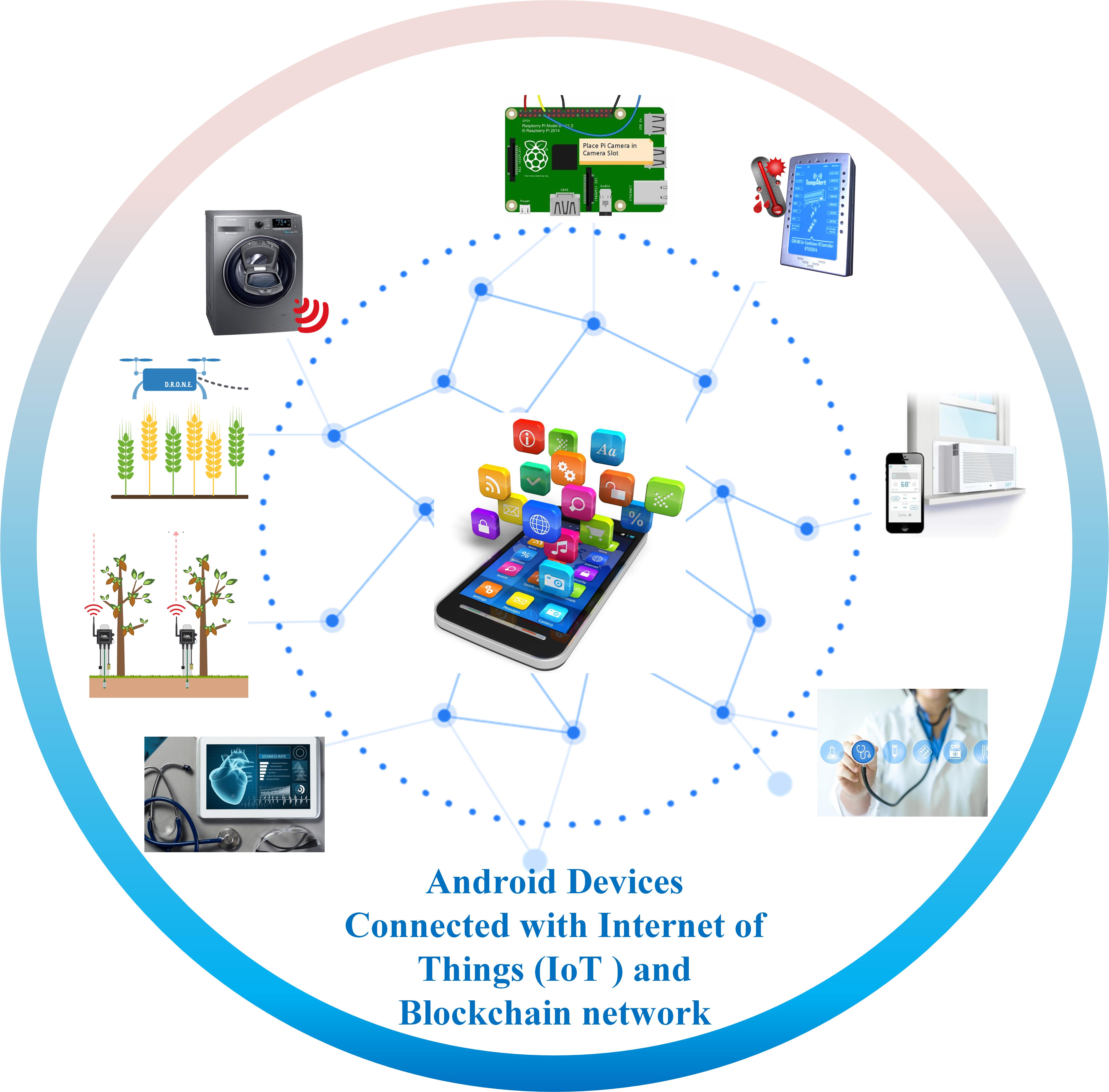}
\caption{An illustration for the integration of IoT smart devices connected on a common network via Android application platform.\label{fig:IoT-devices-Connected}}
\end{figure}

The future wireless technologies such as fifth-generation mobile phone networks (5G) and Internet of Things (IoT) are revolutionizing the world by introducing innovative applications and smart systems that can be only imagined in the past, such as smart environment sensing, smart agriculture, smart drones, smart healthcare monitoring, autonomous cars, and many more. For developing such smart systems, heterogeneous electronic devices participate in a common network to communicate with each other as illustrated in Figure \ref{fig:IoT-devices-Connected}. A wide range of advanced electronic devices are controlled with a powerful Android platform which enables the integration of smart gadgets such as sensors, smartphones, smartwatches, smart washing machines, etc. Such electronic devices including smartphones encourage people to store and share their personal and confidential information. At the same time, it makes these devices become an intensive target for malicious applications to harm users due to common Android platform \cite{Damshenas2015d,DBLP:journals/compsec/CaiLX21,DBLP:journals/compsec/LingSAH21,DBLP:journals/compsec/JerbiDBS20,kumar2018effective,kumar2020}. The attacker exploits the Android system by indulging fake applications that will directly affect the users' privacy and security. It may pose a severe threat by snooping on users' data such as confidential contracts, photos, contact information, location, account information, and passwords. Additionally, the malicious applications can produce adverse effects not only on the intended node but even can affect other linked devices with a shared network. About 0.7 million applications were reported as malicious and blocked by Google Play store before user downloading in the year 2019 \cite{GoogleBlock}. However, most Android app markets do not provide a way to access whether a mobile app is counterfeited or not. Besides, many users install applications from anonymous sources and do not use antivirus applications to protect from malicious and phishing attacks \cite{Walls2015,Dogru2018}. Therefore, there exists an urgent need for an evolved approach and framework that can detect malware applications timely.

Conventional malware detection techniques \cite{Alzaylaee2020,Zhong2019,zhang2019familial,Kim2019,DBLP:journals/compsec/GibertMP20}, formally classified as signature-based \cite{Martin2019,Saracino2018}, access control-based \cite{Sharmeen2018}, and sandbox-based mechanisms\cite{Businesswire2016,Demontis2017,Yerima2016} are significantly dependent on handcrafted features and computationally expensive models. Additionally, most of these techniques are efficient and effective only under some hypothetical constraints which are beyond the real-world scenarios. Recently, deep learning techniques gained significant attention to solve the malware detection problem \cite{Alzaylaee2020,Kim2019,DBLP:journals/compsec/OlukoyaMO20a}. Mainly, previous algorithms are highly based on initial feature extraction processes such as Convolutional Neural Network (CNN), Recurrent Neural Network (RNN), and Long Short-Term Memory (LSTM). However, these techniques cannot be directly applied to smartphones and IoT devices due to their limited resources regarding memory, processing power limitations, and so on. For this purpose, we propose a novel technique to integrate blockchain technology with deep neural networks in order to resolve the limitations of previous malware detection techniques. Our approach enables direct implication for IoT devices.

First, we considered the problem of aggregate and training the learning model with a decentralized network. More precisely, each client initialized  a  multi-layer deep learning model to  compute the learning parameters for inspect the malware using a trained model. The overall training process is accomplished by means of five simple steps: (i) Selection of the important feature using GINI information gain function, (ii) Division of a dataset into different clusters to obtain the fundamental data distribution for a particular group of malware, (iii) Generation of multiple clusters as a sub-tree of each cluster for the huge number of features, (iv) Choosing the best deep learning model for each cluster to classify the malware and benign from the corresponding data distribution, and (v) aggregate the latest weights using the previously trained model. 
The second problem that we prominently addressed was to track the malware or harmful application when users download certain applications from app networks. We used a blockchain-based authentication mechanism to identify the latest malware information which is detected by the deep learning model. We utilized the InterPlanetary File System (IPFS) based storage of the Android applications and hashes of apps in a blockchain ledger. After uploading the Android app in the IPFS, a deep learning model test (benign, malware) app. Finally, the harmful application information store in the blockchain ledger for future identification. In this way, the smart contract helps to approve or deny an application during the uploading and downloading process.

The third problem that we addressed is related to resource consumption in a malware detection model. We combine the local updates from the various clients into a blockchain ledger, once block is mined, it is appended to the blockchain ledger and broadcast latest malware features to the  entire network. The blockchain compute a new version of the global model. The process iterates until global loss function converges or the desired accuracy achieved. In this way, the aggregated deep learning weights and utilize the IPFS technique to reduce the computational cost  and achieve the better intelligent network scalability. The integration of the federated learning model and blockchain collects the new types of malicious features for the Android applications from various sources to train the global model itself. It provides security and makes better detection of malware for Android IoT devices in a real-time environment to protect the potential vulnerabilities attacks. In short, our key contributions can be outlined as follows:

\begin{enumerate}

\item This paper design smart contract which provide a secure downloading and uploading mechanism for Android applications.
	
\item This paper proposes a framework that integrates federated learning and blockchain for better malware detection and information sharing regarding Android applications across the network.

\item The enhancement in a multi-level deep learning model is proposed that can extract multiple types of malware features and the training task is distributed in the blockchain network for better prediction.

\item We performed an extensive empirical analysis to prove the significance of the proposed approach by providing multi-level deep learning and secure data sharing via blockchain.
\end{enumerate}

The remaining part of this paper is organized as follows: Section \ref{sec:related-work} presents the literature review of Android malware detection, and it discusses the static, dynamic, and hybrid analysis. Section \ref{sec:preli} discusses the preliminaries and need analysis for the proposed framework. Section \ref{sec:PModel} states the formulations and work flow of proposed model based on blockchain and deep learning. Next, Section \ref{sec:Perf} analyzes the results and provides a comparison with other works. Finally, we concluded our work in Section \ref{sec:Con}.

\section{RELATED WORK \label{sec:related-work}}
Deep Learning and blockchain are disruptive technologies that changed the landscape of cyber security research. These models have many advantages over traditional machine learning models, especially when there exists a large amount of data. Android malware detection in IoT devices qualifies as a big data problem because of the growing number of malicious applications, the mystification behavior of Android malware, and the potential protection of huge values of data assets stored on the IoT devices. In this section, we introduced a step-wise progress in malware detection for Android  devices. Also, we present a brief overview of blockchain technology for Android IoT security and service enhancements. 

\subsection{Android Devices Security For  IoT }
The Android operating system occupies the worldwide mobile market with billions of devices because of the systems' flexiable nature and accessibility. Every moment, millions of Android apps are ready for end-users for installation through a variety of app stores such as Google Play. Unfortunately, Android attackers take advantage of growing Android apps and spread malicious applications with an over 10 times increment of the number of detected Android malware reported between 2012 and 2018 \cite{DBLP:journals/csur/QiuZLPNX21}. Furthermore, according to Google officals, over 12K fresh Android malware samples were encountered daily in 2018. In addition to the existing number of growing attacks, newly released Android malware samples are more sophisticated than the samples that appeared in the past.

With the popularity of IoT devices in digital-world and widespread IoT platforms, users personal information stored in IoT networks got a considerable amount of attackers attention in recent years. To achieve a secure, robust, and reliable communication among IoT devices equipped with camera and microphone, like smartphones, researchers \cite{zhou2018dolphin, zhou2018enabling} utilize voice and visible light to transmit data among IoT devices. For example, Zou et al. \cite{zou2017robust} proposed a model to identify a specific individual with his unique gait feature obtained by using RGBD sensors deployed in the house to prevent burglary. A considerable amount of research has been published for the security of IoT environment by designing access control system. For example, Luo et al. \cite{ DBLP:journals/iotj/LuoCHPY21} proposed contextual model for privacy and security through inferring apps in smart home. Also, Mossain et al. \cite{DBLP:journals/tccn/HossainX20} introduced a context-aware framework for detection of hidden terminal emulation attacks in cognitive radio-enabled IoT networks. In the same stream Jia et al. \cite{jia2017contexlot} introduced a context-based permission system for IoT platforms that provides contextual integrity and runtime prompts. 
Kousalya et al. proposed a reliable service availability and access control method for cloud assisted IoT devices. The proposed method is robust for inherent synchronization issues, resource availability and accessibility difficulties. Additionally, Pallavi and Kumar \cite{DBLP:journals/wpc/PallaviK21} take advantage of fog computing to secure IoT devices by considering a trusted third party authentication scheme. The authentication scheme using fog node offers reliable verification between the data owners and the requester without depending on the arbitrators' trust. The proposed scheme also, effectively resolve the problems related to the single point of failure in the storage system and offers significant benefits by increasing the throughput and reducing the computation cost. Lee and Lee \cite{DBLP:journals/mis/LeeL21} introduced state-of-the-art research trends and recent developments in IoT security. A plenty of other researchers \cite{DBLP:journals/access/LiaoANHK20,DBLP:journals/cluster/AlfandiKAK21,DBLP:journals/fi/YuZCSJ20,DBLP:journals/iotj/MeneghelloCZPZ19,DBLP:journals/istr/AmmarRC18} highlighted significance of malware detection and IoT security for Android devices. 

In parallel with machine learning algorithms, the researchers have also started using innovative blockchain technique to protect the underlying IoT smart devices \cite{StopTheFakes2018,Gu2018,BlockVerify2019,Alzahrani2018,kumar2021integration,kumar2020blockchain,DBLP:journals/cluster/AlfandiKAK21}. Blockchain, often confused by some as a synonym to Bitcoin, is the technology behind this in famous cryptocurrency. It is a distributed ledger that stores the data in blocks. These blocks are in order and linked with each other cryptographically forming a chain in a way that makes it computationally infeasible to alter the data in a particular block \cite{Alzahrani2018}. This mechanism ensures transparency, decentralization, verifiability, fault-tolerance, audit-ability, and trust \cite{Alzahrani2018,BlockVerify2019}. There is no single consensus on the types blockchain, most commonly the blockchain techniques are distributed as public, private, and consortium. Public or permission-less blockchain technique and is open to everyone, so anyone can access and use them. On the other hand, private or permissioned blockchain techniques are controlled by one or few specified authorities, hence, not everyone can access them. The transactions here are faster and only the selected few are authorized to approve a transaction consensus. A plenty of researchers highlighted inbuilt power of blockchain to protect IoT smart devices \cite{DBLP:journals/fi/YuZCSJ20,StopTheFakes2018,Gu2018,BlockVerify2019,Alzahrani2018,kumar2021integration,kumar2020blockchain,DBLP:journals/cluster/AlfandiKAK21}. Although, certain machine learning frameworks and blockchain based techniques have been developed to deal with cyber threats in the IoT domain, combining these two is something new that needs to be explored.

\subsection{IoT Platform}
IoT platforms are used to build applications that monitor and control the IoT devices. These platforms provide developers with the ability to quickly build, test, deploy, and iterate on IoT-specific applications. According to a technical report by G2 rating database, the global market share of IoT platforms is going to reach till \$ 74.74 billion by 2023. The reason behind this growth is the huge demand for IoT devices and other components. Therefore, leading technology stakeholders struggling to win the race of providing sustainable IoT platforms. As per current market share, the most popular IoT platforms includes Google Cloud IoT Core, IBM Watson IoT Platform, AWS IoT Core, Particle, Microsoft Azure IoT Suite, Oracle IoT platform, and many more. Also, these platforms act as a central hub where other smart devices interact with each other and a cloud is used to synchronize device states. These devices collect physical information and send events to the cloud to trigger other events. Mostly, applications for IoT platforms are developed in Groovy language, which is based on JVM, and executed in a Kohsuke sandboxed environment. Compared with Android applications, these are simpler and can be processed in the similar way with Android applications.

\section{PRELIMINARIES \label{sec:preli}}

This section quickly review the some preliminaries about
Android application analysis techniques. We divide this
section into four parts i) Static analysis, which contains two approaches
the first approach is permission-based, and the second approach is
API Call. ii) Dynamic analysis that is used to extract real-time phone
features, and iii) Hybrid analysis that combines the static and dynamic
features. Final part provides a comparison between all techniques.

\subsection{Static Analysis}

Static analysis can check the application\textquoteright s behavior
without executing the app. Several machine learning techinques are
proposed to classify benign apps and malicious apps\textbf{ }\cite{Yerima2016,Yerima2015,taheri2020similarity}
such as content based analysis that reduce the dimensions of the content.
The latest research \cite{DBLP:journals/compsec/HanXWHKM19,Arp2014,Karbab2018,Kumar2019}
for static analysis based on~the API calls and permission features. 
The malware detection is less effective is such classical techniques
\cite{Westyarian2015,wang2020deep,zhu2020sedmdroid,DBLP:journals/compsec/LiuLLZ20,DBLP:journals/compsec/VermaMS20,DBLP:journals/compsec/JerbiDBS20,DBLP:journals/compsec/UcciAB19,feng2020performance,DBLP:journals/compsec/HanXWHKM19}.
However, low efficiency is demonstrated using these methods for feature
extraction. Our main focus to design a multi-level deep learning model,
which can support a various kind of features and classify the malware
and benign effectively.

\begin{figure}
\includegraphics[scale=0.3]{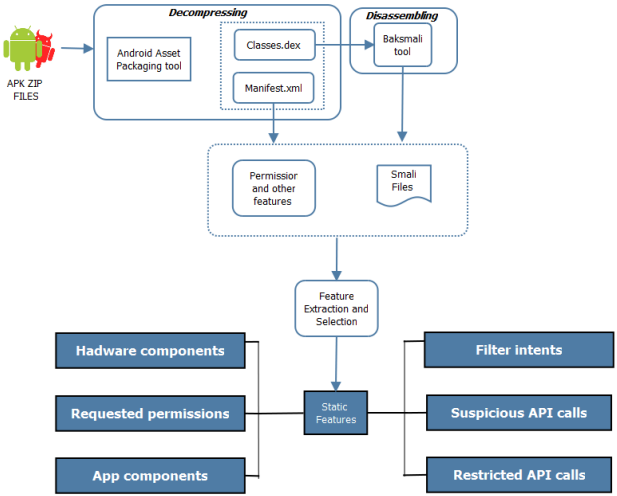}\caption{Static feature extraction process\label{fig:Static-Feature-Extraction}}
\end{figure}

\subsubsection{Permission-based analysis}

Android uses a permission-based security model to ensure that sensitive
information of the user is restricted, and the actual user only can
access it. Indeed, permission is the most effective static feature
because attackers apply for permission to reach their malicious goals.
Before the app gets installed, it asks for some requested permissions
from the user. After permission granted, the app installs itself on
the device. There are many approaches that extract permissions for
malware detection \cite{DBLP:journals/compsec/BlackGL18,Wang2014,Kumar2019,Li2018d}.
Wang et al. \cite{Wang2014} proposed a methodology for analyzing
the permission-based on permission ranking, association rule and similarly
based . It finds the permission groups using the correlation coefficient
and ranks each permission individually. Verma et al.,\cite{Yerima2014,Yerima2015,Yerima2016}
used the information gain algorithm of feature selection to choose
the best features from android apk packed files.

\subsubsection{API calls}
API stands for Application Program Interface. API calls are used by
the apps to interact with the Android framework. Some works target
API calls and use it as a promised feature to investigate malicious
behaviour.

\subsection{Dynamic Analysis}

Moreover, dynamic analysis \cite{Zhao2011} was proposed to observe
to real-time behavior of the phone to observe the dynamic behaviors
and features of applications. In this perspective, to analysis, the
dynamic behavior of malware activities, use the Emulator (Android
Virtual Device) to extract the dynamic features such as API calls,
Events/Action. There are several tools to use dynamic analysis methods,
including the Monkey and the DroidBot tool. Using these tools, Dynalog
is an essential file used to input generation methods. It can extract
the features of the API calls and system call information that reveals
how malware behaves, the features of the dynamic methodology can be
observed in \cite{Isohara2011,Burguera2011,Ham2014,Ham2014a,Huang2013}
and detection of unknown malware that shows similar behavior is also
possible using these methods \cite{Ham2014,Damshenas2015d,Sun2017a}.
Also, the API call analysis and control flow are the dynamic analysis
methods \cite{Li2018d,Zhao2011,Wu2014,Afonso2015}. The main difference
between existing works and ours is that our approach combines both
static and dynamic analysis methods with the blockchain \cite{kumar2020blockchain,kumar2021integration} and deep learning
to increase the detection rate and overcome machine learning weakness.
Furthermore, our approach is to distribute the malware information
in the blockchain network that can notify benign and malware Android
apps at the installation time.

\begin{figure}
\includegraphics[scale=0.17]{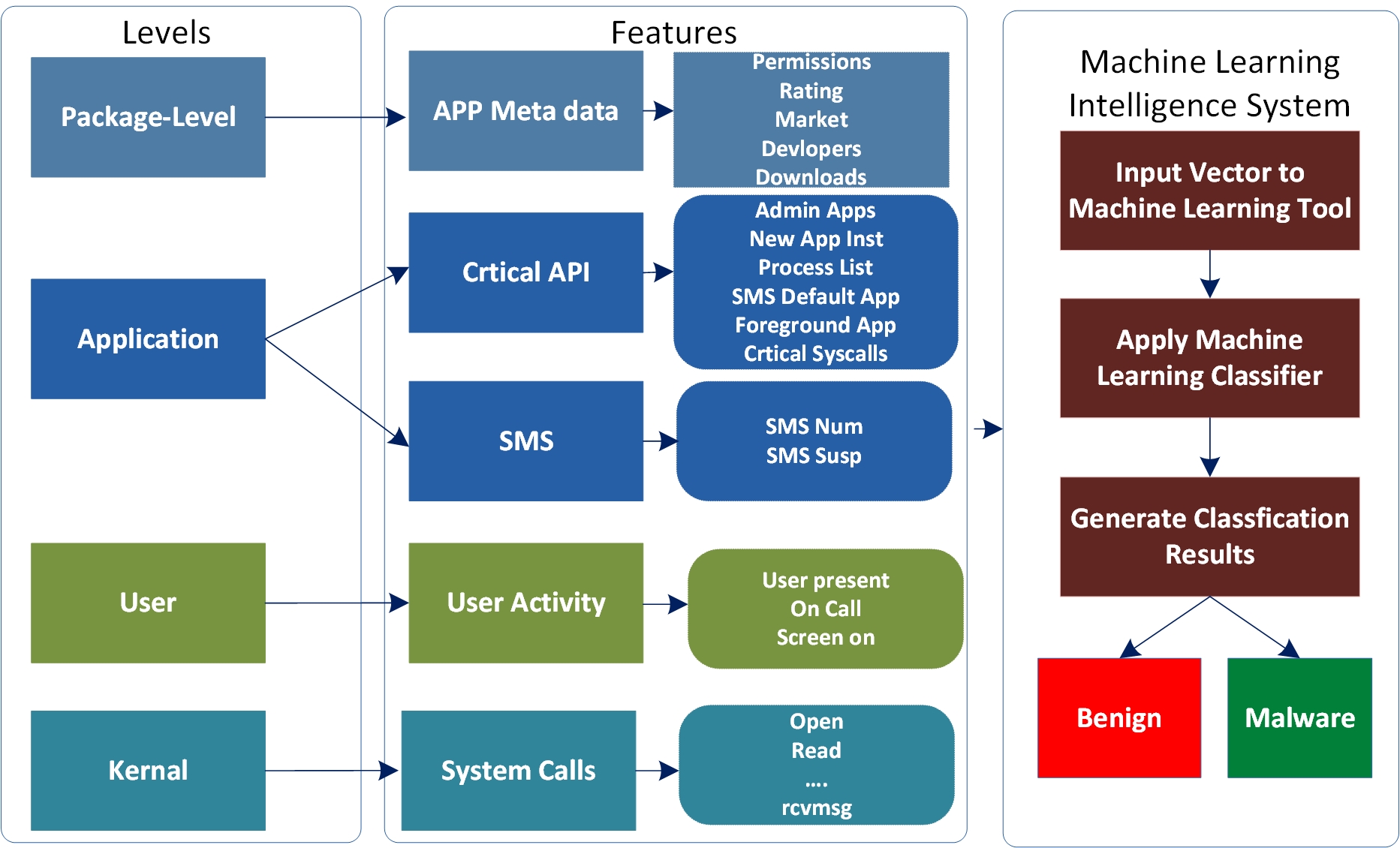}\caption{Traditional dynamic analysis features extraction and classification
process \label{fig:Dynamic-Analysis}}
\end{figure}

\subsection{Hybrid Analysis}

\begin{figure}
\includegraphics[scale=0.32]{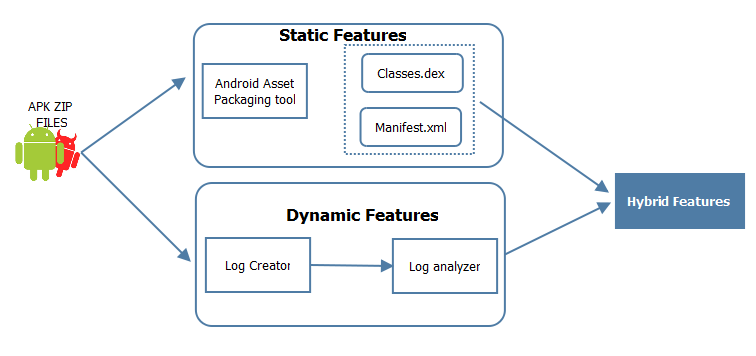}\caption{Hybrid feature extraction process \label{fig:Hybrid Analysis}}
\end{figure}

The hybrid analysis combines static analysis and dynamic analysis.
The static features are extracted without executing the application.
In contrast, dynamic features are extracted by an emulator or on the
real device, which is time and resource-consuming. The hybrid analysis
illustrates in Figure \ref{fig:Hybrid Analysis} to combine the static
and dynamic analysis. Some researchers focus on hybrid analysis \cite{Ferrante2018,Kim2013,Saracino2018,Huda2018a,Liu2016},
but these methods are time-consuming. To solve this problem, we design
the blockchain-based framework that equally distributed resources
to all users. Our proposed technique is more sufficient and less time
consuming because of the distributed nature of blockchain.

\section{SYSTEM MODEL \label{sec:PModel}}

\subsection{Proposed Architecture}
In this article, we consider the problem of sharing the latest malware
features and train the deep learning model in the decentralized network
for multiple Android IoT devices. Due to limited resources and power
consumption of Android devices. We design a multi-feature deep learning
model which support various features form the decentralized network.
Then we are focusing the aggregate a previously trained model with
new features sets from the latest apps which are uploaded recently.
The harmful application information stores in the blockchain network.
Finally, we provide security of the android derives using a smart
contract for retrieving the information of the malicious Android applications.
Figure \ref{fig:layer-blockchain} shows the architecture of the proposed
framework. The steps of proposed framework shown in below:
\begin{figure*}
	\centering
	
	\includegraphics[scale=0.35]{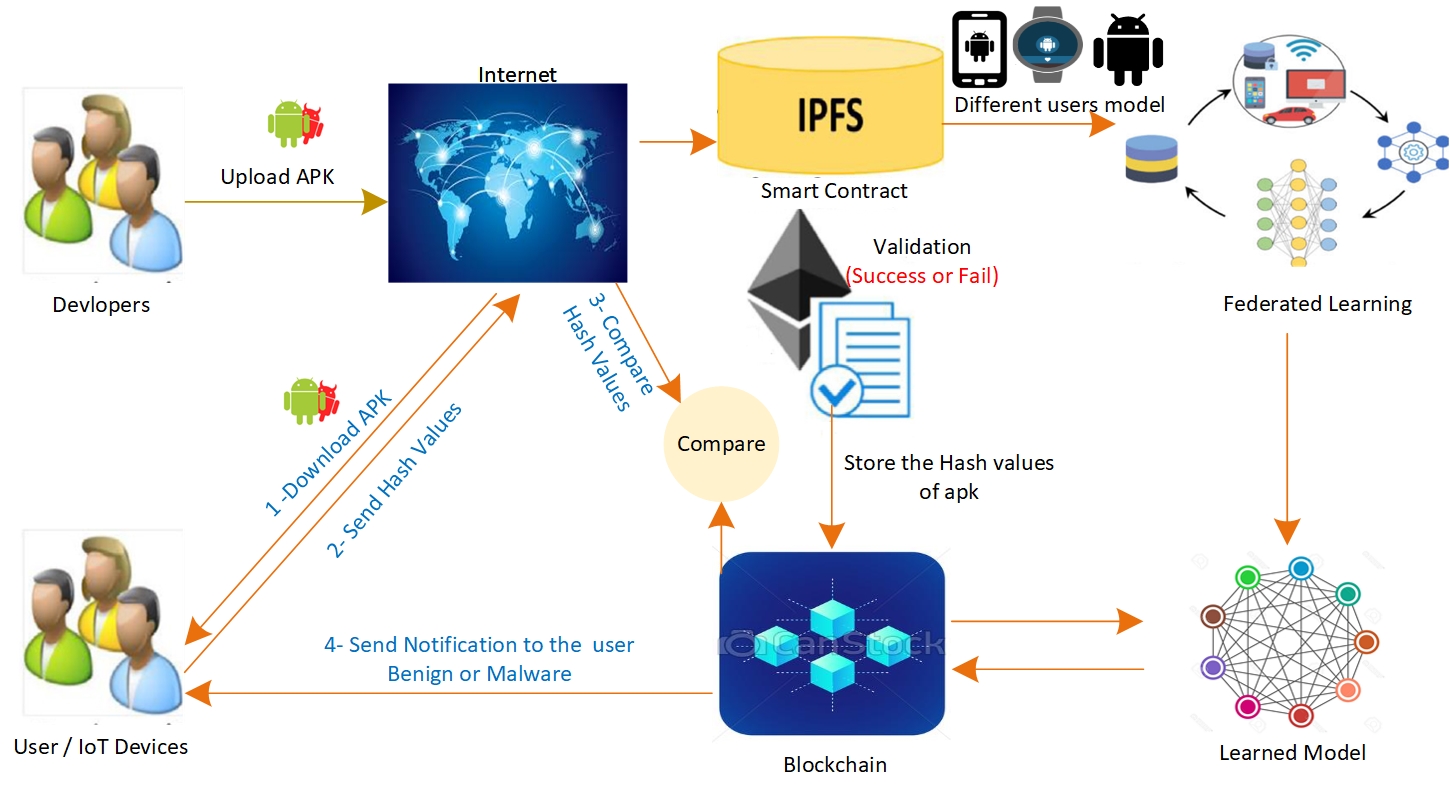}
	
	\caption{Proposed framework based on Federated learning and blockchain}
	\label{fig:layer-blockchain}
\end{figure*}
\begin{figure}
	\includegraphics[scale=0.08]{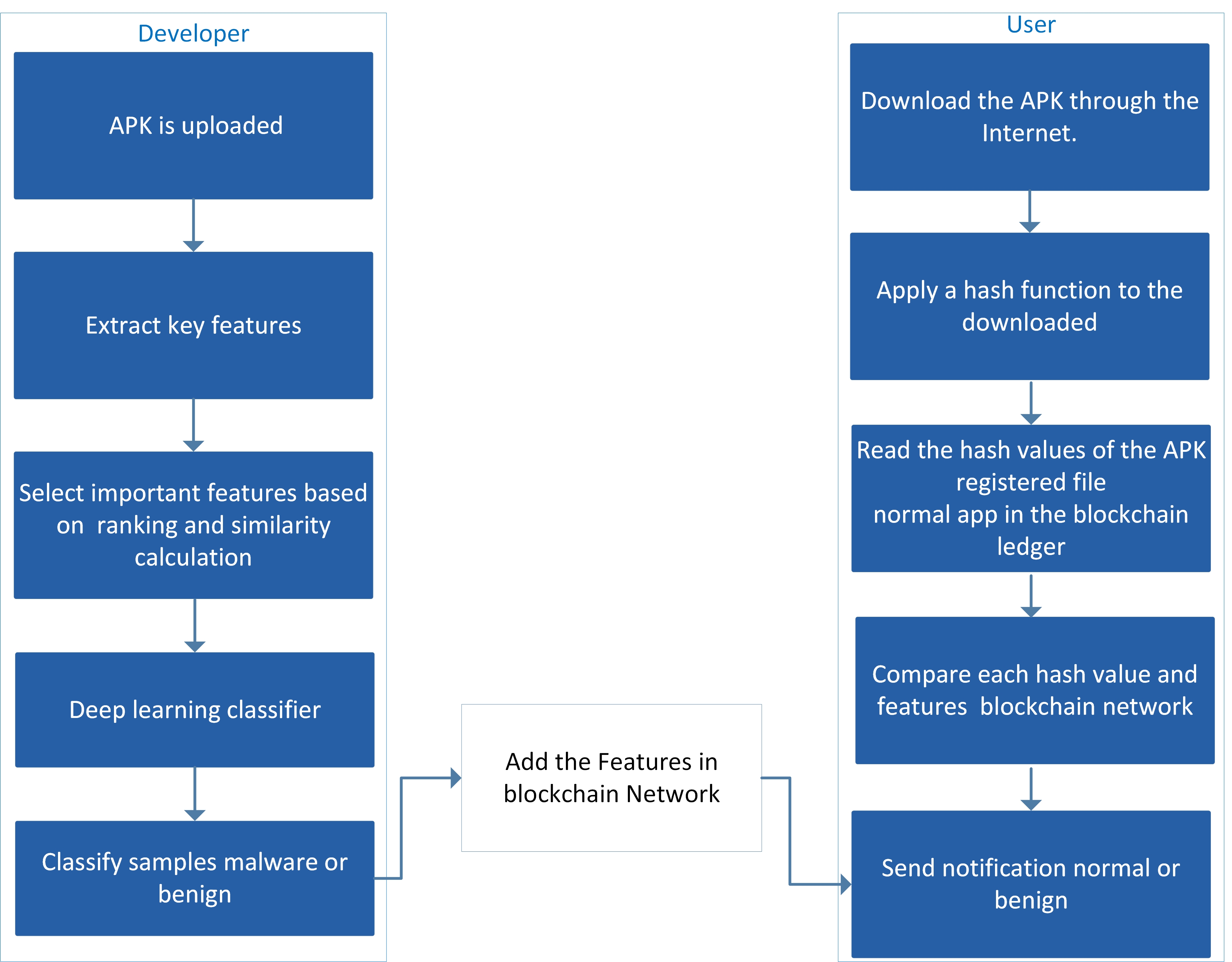}\caption{Flowchart of user and developer for Android malware detection using blockchain
		\label{fig:layer-blockchain-3}}
\end{figure}

\begin{enumerate}
\item The user uploads app to the network
\item The version of the app will be stored in the InterPlanetary File System
(IPFS) .
\item The deep learning model extracts the benign and malware features from
the uploaded app. More detail in section \ref{subsec:Probability-Based-ProdDroid-1}
.
\item The updated deep learning model store in IPFS system for reducing
the cost of blockchain. Additionally, we aggregate the model weights
to reduce the computational task in the blockchain nodes. In section
\ref{subsec:Aggregate-deep-learning}
\item The hash value of app and decision results obtained from deep learning
model is stored in a distributed ledger. More details shown in section
\ref{subsec:Storing-information-about}
\item During the downloading process of the app: i) the user will send hash
value to the network ii) the smart contract will come into action
to compare and verify the hash value of the downloaded app iii) Finally,
it will notify the user regarding the malicious or benign app. More
details shown in section \ref{subsec:smart-contract}
\end{enumerate}
In summarizing Figure 8 provide a more detailed description of uploading
or downloading from a developer and user perspective. When a user
downloads the Android apps or developers uploading the apps, the smart
contract uses for secure data uploading and check the harmful features
of the apps automatically. The smart contracts can track malicious
apps from the decentralized network. Also, it can help to learn the
deep learning model itself and track the malware application when
users are downloading the app through the Internet. Additionally,
blockchain technology can create a trust-less environment and guarantees
the transparency and reliability of the distributed nodes.

\subsection{Proposed Deep learning local model  \label{subsec:Probability-Based-ProdDroid-1}}

\begin{figure}
\centering

\includegraphics[scale=0.15]{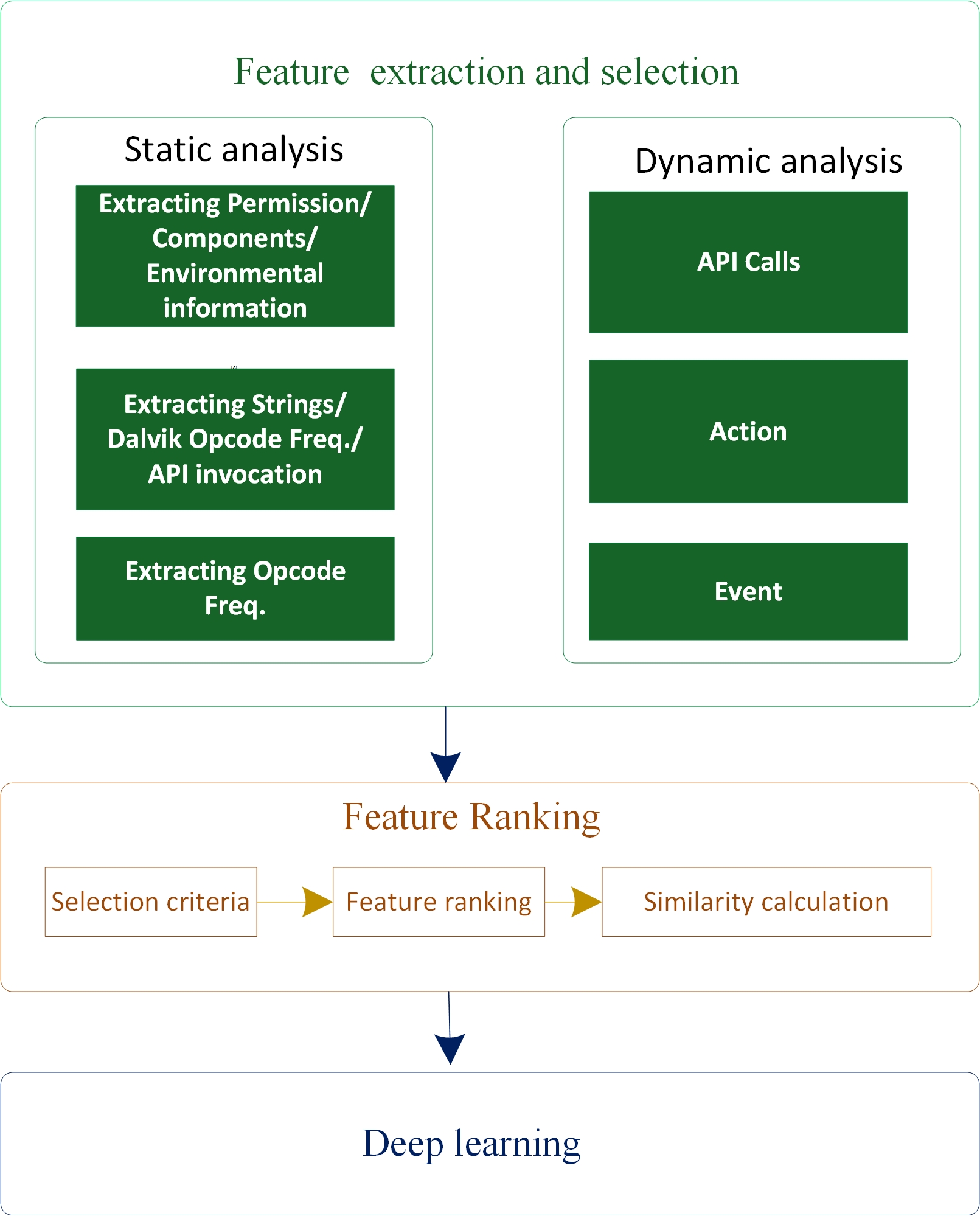}

\caption{Static and dynamic analysis local deep learning model }
\label{fig:Proposed-architecture}
\end{figure}

In this section, we proposed a deep learning-based model based on
static and dynamic analysis. Figure \ref{fig:Proposed-architecture}
shows the overall architecture of the deep learning model for static
and dynamic analysis. In the first phase, we combine static and dynamic
features. The static features are extracted by decompiling the Android
apk file, and dynamic features are gathered from Droid Emulator(Android
Virtual Environment). The DroidBot Emulator generates the Dynalog
file, and for static features, we use CSV file. In the second phase,
select the static and dynamic features from the CSV and Dynalog file
and rank the features using information gain function. The information
gain function score the features such as TelephonyManager;- > getDeviceId
is 0.98. After that, we found the similarity among the features. In
the third phase, deep learning evaluates the performance of benign
and malware applications and train the classifier. In the fourth phase,
share the features information of malware and benign in the blockchain
distributed database for achieving real-time malware detection for
Android IoT devices.

\subsubsection{Feature selection for hybrid malware detection}

We used the feature importance property of the model. Feature importance
gives a score for each feature of data between zero and one. The higher
the score is, the more important or relevant is the feature towards
the output variable. This score helps in choosing the most important
features and drop the least important ones for model building. Feature
importance is an inbuilt class that comes with tree-based classifiers.
The information gain (IG) was used to select important features with
a high score to classify the data effectively\cite{Han2012}. The
information gain express in the \ref{eq:1}, \ref{eq:2}, and \ref{eq:3}.equation

\begin{equation}
{Gain}(A)={Info}(D)-{Info}_{A}(D)\label{eq:1}
\end{equation}

\begin{equation}
info(D)=-\frac{pos}{total}\log_{2}\frac{pos}{total}-\frac{neg}{total}\log_{2}\frac{neg}{total}\label{eq:2}
\end{equation}

\begin{equation}
\text{ InfoGainRatio }(A)=\frac{Gain(A)}{Info(D)}\label{eq:3}
\end{equation}

\subsubsection{Deep learning model training}

\begin{figure}[h]
	\centering
\includegraphics[scale=0.11]{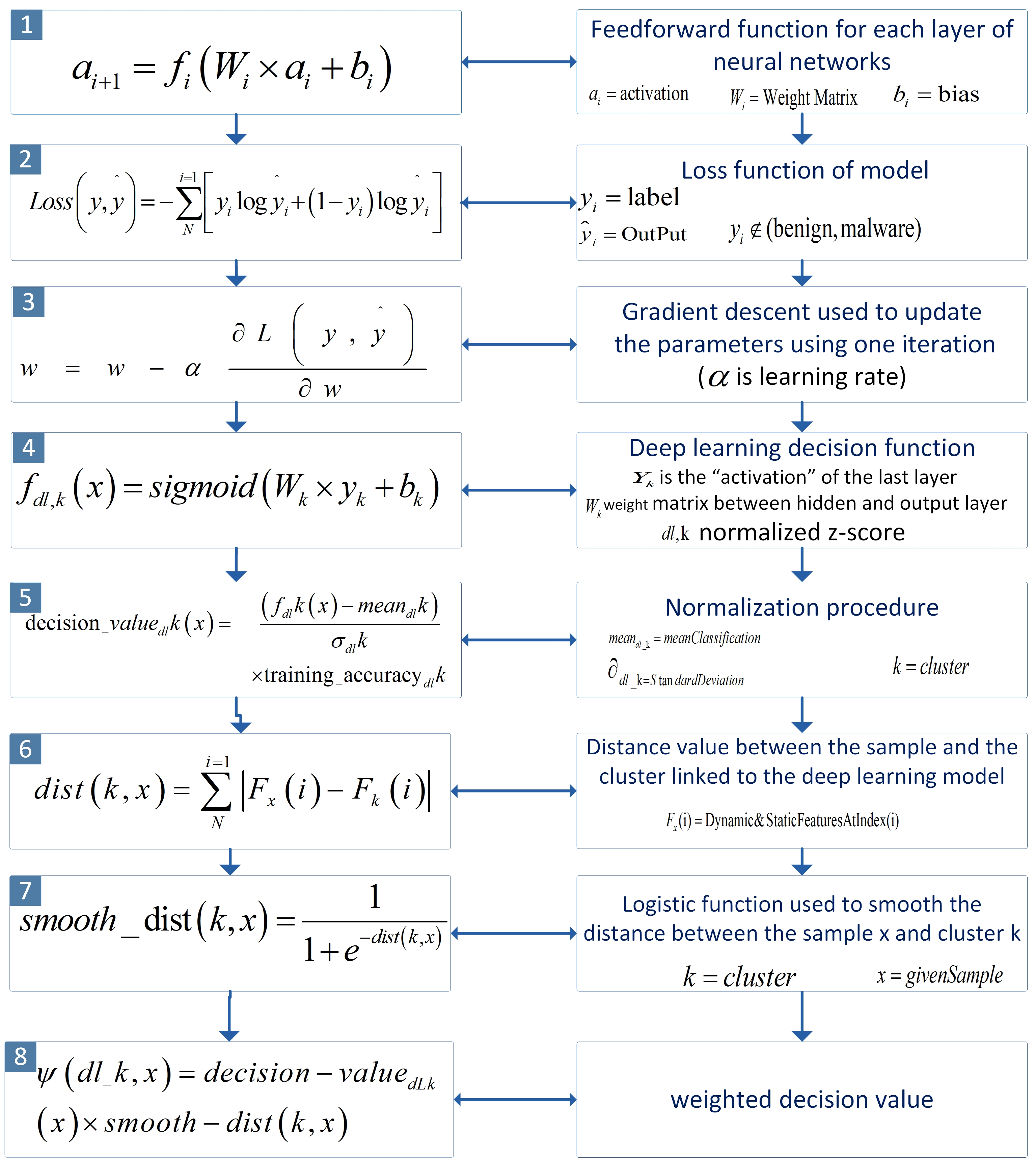}\caption{Deep learning model training steps.\label{fig:Deep-Learning-Modal}}
\end{figure}

\begin{figure}[h]
\includegraphics[scale=0.11]{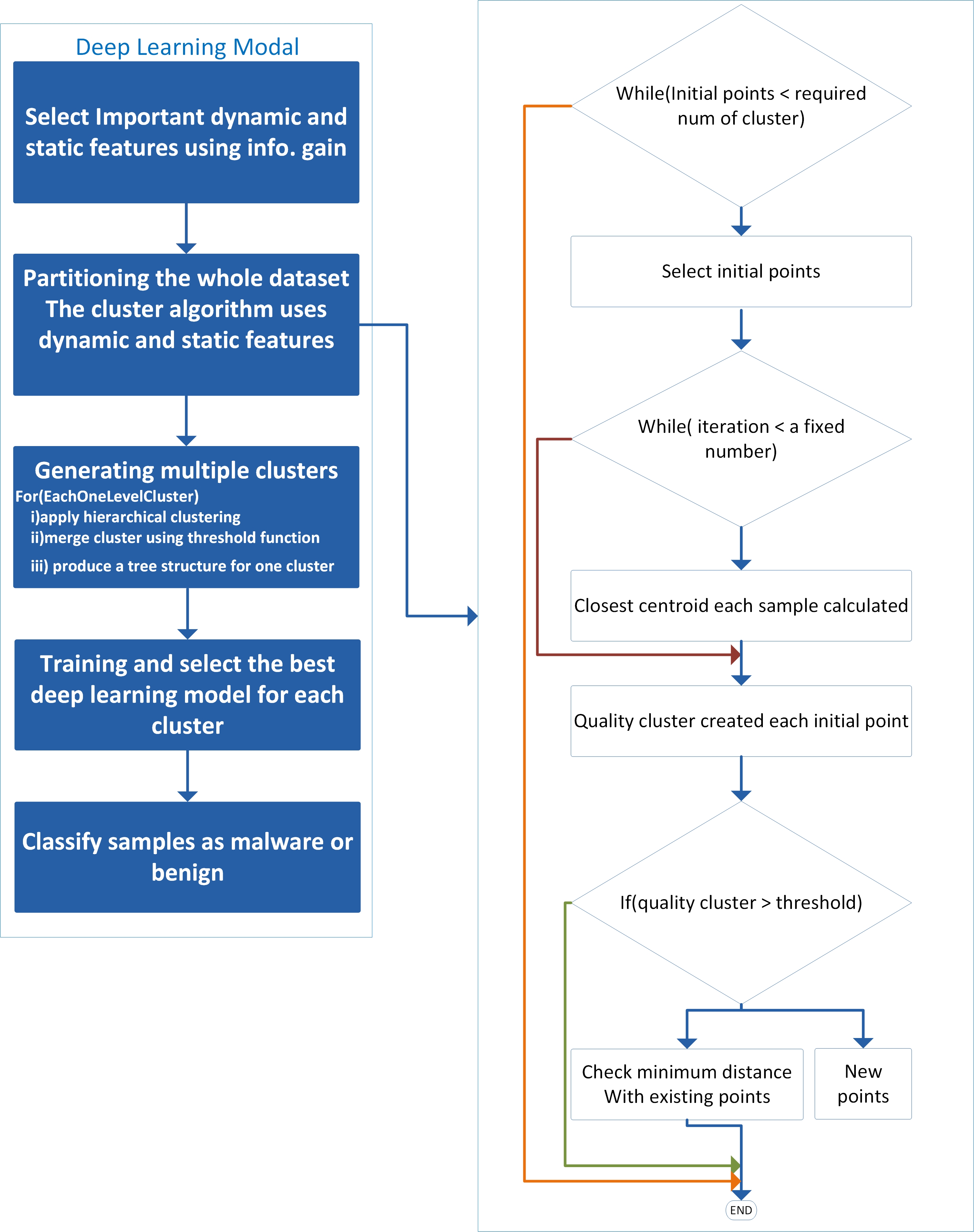}\caption{Proposed Android multi-feature deep learning model.\label{fig:Proposed-Android-Multi-Feature}}
\end{figure}

Our main aim is to create a deep learning based model that ensures
that Android malware and benign are accurately classified. Furthermore,
it detects Android malware from benign apps. The previous paper discusses
various techniques for malware detection \cite{Alzaylaee2020,Zhong2019,Nguyen2018,Kim2019,DBLP:journals/compsec/GibertMP20}.
These methods are highly efficient, though; however, they can not
be applied directly to mobile and the IoT devices. To improve the
detection performance of deep learning model, we design the multiple-level
of deep learning. Each deep learning model learns from specific features
of Android malware data for a single group of malware. Finally, all
groups of deep learning models combined and make final predication.
We tested some deep learning models during the training process, which
include deep recurrent neural Network, Convolutional Neural Network,
Fully Feed Forward Network. One of then us the best for each cluster.
Additionally, LSTM avoids the batch normalization vanishing gradient
problem for multiple features. We combine the RNN and LSTM to achieve
better performance in distinguishing the malware and benign application.
In the first step, we select the important feature using information
gain function for the static and dynamic analysis. In the second stage,
the dataset dived into different clusters that calculate the unique
data distribution. In the third stage, multiple clusters generated
as a sub-tree of each tree cluster for the huge number of features.
In the fourth stage, the best deep learning classifier is selected
to distribute the malware and benign from the unique data distribution
for every cluster. However, the proposed deep learning model classifies
every cluster of each distinct feature for the static and dynamic
analysis. The use of multiple deep learning models during the training
phase reduces time and provide better accuracy. Finally, our proposed
model classifies the malware and benign. The use of multiple deep
learning models during the training phase reduces time and provides
better efficiency. Finally, our proposed model classifies the malware
and benign. The workflow of all stages shown in Figure \ref{fig:Proposed-Android-Multi-Feature},
and deep learning model training is shown in Figure \ref{fig:Deep-Learning-Modal}.
Therefore, our model improves the detection performance and efficiency
of the traditional deep learning classifier.

Moreover, to save the computational power, the training task is distributed
in the blockchain network through the forward propagation and backward
propagation. In the forward propagation, the input is passed through
the blockchain decentralized network, and after processing the input,
the output is shared in the decentralized network. Then, in backward
propagation, the weights of neural networks are shared to the blockchain
network to reduce the computational power. Therefore, the distribution
of the training task reduces time and utilized decentralized resources
over the network. Additionally, we simulate the training outputs in
the decentralized network through the Proof-of-Work, Proof-of-Stake
and Delegated-Proof-of-Stake. Proof-of-Work reduces the computational
power of the deep learning model. Delegated-Proof-of-Stake is using
to vote the hash. it avoids the complex hash operation. More precisely,
state information of each node in a distributed network is taken as
the dataset. The dimensional matrix $(M)$ is the input of the deep
neural network, and the average number of becoming the mining node
in a term is the capacity label. After training our network, we can
get the average transaction number of the $i_{th}$ node as long as
$M_{i}$ is input. Finally, implementation of the deep neural network
is aimed to learn itself from the huge volumes of data resources through
the blockchain technology. The next section discusses the blockchain
technology.

\subsection{Proposed blockchain based Federated Learning model \label{subsec:Aggregate-deep-learning}}

This section aggregates the local trained model with the new
information of the latest feature application features and updated
the latest model in the IPFS to track the new harmful apps. The process
of combining the blockchain and federated learning technology shown in
Algorithm 1. Neural networks are trained through (i) forward propagation (ii) backward
propagation are considered to calculate layers\textquoteright{} weights.
In the forward propagation, the input is passed through $F=f(x,w)=\bar{y},$
and to processing the code $x$ is input and $w$ parameter vector,
the trained malware and benign features set $F={(x_{i},y_{i});i\epsilon I}$
for each devices $(x_{i},y_{i})$. The output weights are shared in
the decentralized network through the IPFS. The loss function of the
training feature set is defined as $L(F,w)=loss$. $F$ is defined
as dataset and $l$ is the loss function. $loss=\frac{1}{F}\sum_{\left(\mathbf{x}_{i},\mathbf{y}_{i}\right)\in F}l\left(\mathbf{y}_{i},f\left(\mathbf{x}_{i},\mathbf{w}\right)\right)$
Then, in backward propagation, the updated weights of neural networks
are using stochastic gradient descent (SGD) defined as below equation:

\begin{equation}
\mathbf{w}^{t+1}\leftarrow\mathbf{w}^{t}-\eta\nabla_{\mathbf{w}}L\left(F^{t},\mathbf{w}^{t}\right)\label{eq:b2}
\end{equation}

As we can see in equation \ref{eq:b2} , the learning rate is $\eta$
, and the $i^{th}$ is the iteration of the $\mathbf{w}^{t}$ parameters.
$F^{t}\subseteq F$ is the each devises mini-batch training dataset.

The above equation is use for single user. Moreover, to learn local
model collaboratively and create a global model from the every devices
$v\in V$ shown in equation \ref{eq:b3}

\begin{equation}
\mathbf{w}^{t+1}\leftarrow\mathbf{w}^{t}-\eta\frac{\sum_{v\in V}\nabla_{\mathbf{w}}L\left(F_{v}^{t},\mathbf{w}^{t}\right)}{|V|}\label{eq:b3}
\end{equation}

\begin{algorithm}
\SetAlgoLined

MD $\gets$ MobileDevices \;
 $\left\{F_{n}\right\}^{n \in[N]}$  $\gets$ Malware Features   \;
 $w^{0}$   $\gets$ global weights \;
$L(w,x)$  $\gets$  Loss \;
$I$  $\gets$  iteration \;
$\theta$  $\gets$ clip bound  \;

\For{i $\in[I]$}{
\For{md $\in[MD]$}{
sample malware and benign feature data set with probability $\frac{\left|f_{md}^{i}\right|}{\left|f_{md}\right|}$ \;
}
\For{$\mathbf{x} \in F_{h}^{t}$}{

$g f_{md}^{i}(\mathbf{x}) \leftarrow \nabla_{\mathbf{w}^{i}} L\left(\mathbf{w}^{i}, \mathbf{x}\right)$ \;
$g f_{md}^{i}(\mathbf{x}) \leftarrow g d_{md}^{i}(\mathbf{x}) / \max \left(1, \frac{\left\|g d_{i}^{2}(\mathbf{x})\right\|}{\theta}\right)$ \;
retrieves  the weights or global model  from permissioned blockchain  \;

}
$g f_{md}^{i} \leftarrow \sum_{\mathbf{x} \in D_{md}^{i}} g d_{md}^{i}(\mathbf{x})+\mathcal{MD}\left(0, \frac{\theta^{2} \rho^{2}}{MD}\right)$\;
executes IPFS model to  aggregation and obtain updated the IPFS model \;
add the parameters of model as a transaction \;
}

$g f_{md}^{i} \leftarrow \frac{1}{MD}\left(\sum_{n \in[md]} g f_{md}^{i}\right)$ \;
$\mathbf{w}^{i+1} \leftarrow \mathbf{w}^{i}-\eta \cdot g d^{i}$ \;
retrieves the current updated weights from IPFS, and aggregates the weights\;
broadcasts new malware information to other delegates for verification, and collects all transactions into a new block\;
appends the block including the global model to the permissioned blockchain\;

\caption{Aggregate deep learning weights from the blockchain network  }
\end{algorithm}


In this way the federated learning model identifies the malware app through compute the gradients and send the updated weights to the global blockchain network. The smart contract shares
the aggregated updated results. Moreover, when the user downloads
the application the smart contract identify the harmful app.

\subsection{Storing information about malware features in blockchain} \label{subsec:Storing-information-about}

\begin{figure}[h]
\includegraphics[scale=0.35]{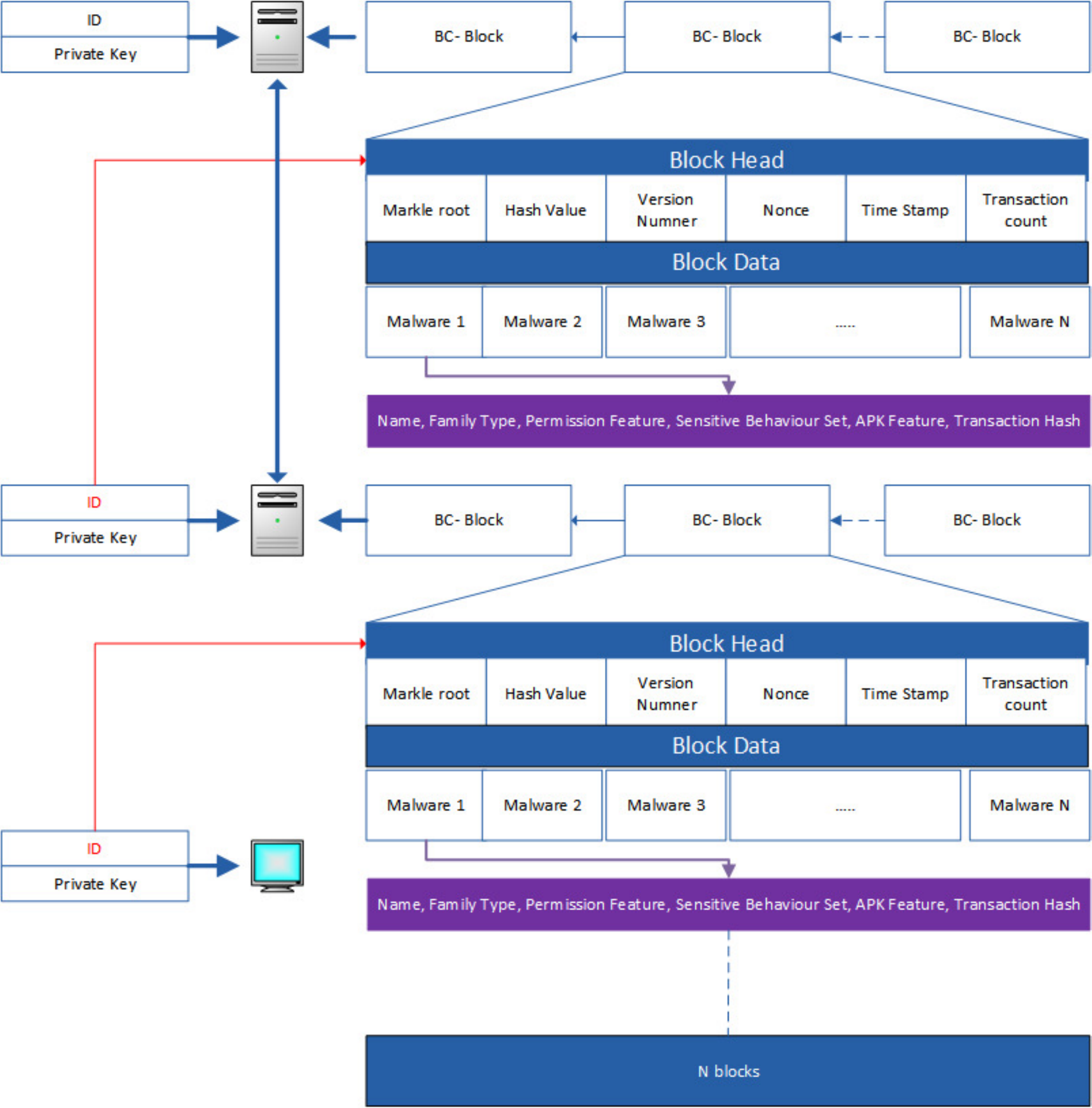}\caption{Blockchain data-store technique for multi-features of Android malware
\label{fig:Blockchain-data-store-techique}}
\end{figure}

We store the Android application hashes with the malicious and benign
features (static and dynamic) in the blockchain distributed database.
The structure of the storing information about malware features shown
in Figure \ref{fig:Blockchain-data-store-techique} and further describes
the attributes of the Figure \ref{fig:Blockchain-data-store-techique}
in Table \ref{tab:Blockchain-Attributes}. Furthermore, the blockchain
structure divided into two parts i) Block header and ii) Block data.
In the first part of block header stores the version number of apps,
Markle root, hash values of all apps, and so on. The second part block
data stores the all static and dynamic features such as suspicious
API, permission, events , calls, etc. The primary purpose to store
the malware information in the blockchain distributed database is
to ensure the security the identical hash values which can effectively
prevent fraud such as de-compile and repacking Android applications
by reverse engineering techniques. Therefore, no one can easily create
counterfeit applications.

Moreover, an existing system such as VirusTotal has flaws in detecting
fake Android mobile apps. The proposed blockchain framework we offer
to remove these flaws and recognize Android fake/malicious applications.
Furthermore, the blockchain includes actual information in a decentralized
malware blockchain database to increase the prediction performance
of the malware and run-time detection of malware when the user downloads
and upload the Android app into the network.

\begin{table}
\caption{Blockchain Attributes \label{tab:Blockchain-Attributes}}

\begin{tabular}{|c|c|>{\centering}p{0.5\columnwidth}|}
\hline
Keywords & Size & Definition\tabularnewline
\hline
\hline
Pre- Hash & 32 bytes & preceding block hash value\tabularnewline
\hline
Version number & 4 bytes & track the protocol or software updates\tabularnewline
\hline
Timestamp & 5 bytes & records the time a block\tabularnewline
\hline
Transaction\_count & 15 bytes & number of malware results in the current block\tabularnewline
\hline
Merkle root & 32 bytes & it calculate the malicious codes which detected by block\tabularnewline
\hline
Nonce & 15 byte & randomly recognized as a formal block\tabularnewline
\hline
\end{tabular}
\end{table}

\subsection{Designing a Smart Contract to secure the Android devices to check
the harmful apps\label{subsec:smart-contract}}

This section describes the utilization of Ethererum blockchain and
smart contract for verification, tracking of versioning history of
Android apps, and further discusses the storing mechanism of hash
values in a distributed ledger. The proposed smart contract can track
different versions of apps and can provide continuous detection by
broadcasting and sharing information regarding every new malicious
app. To store an app on the blockchain network will be very expensive
and wasteful in terms of resources as most apps have relatively large
sizes ranging from several megabytes (MB). That\textquoteright s why,
firstly, the uploaded app by a developer will be stored in the IPFS
file system along with its version history, and further only corresponding
hash values of apk file will be stored in blockchain distributed ledger.
Moreover, the use of the IPFS also provides several other benefits
due to its peer-to-peer network feature and support regarding the
tracking of the versioning history of every uploaded apk files. Furthermore,
our design smart contract interact during the uploading and downloading
Android applications. It handles the Android application IPFS version
and the hash value of the application. It can approve or deny to upload
harmful Android applications during uploading/downloading. Finally,
the malicious features are broadcast using smart contracts to all
users across the network. Figure 4 interacts between participating
entities that are defined as developers, users, approves, deep learning,
and the smart contract.

\textbf{Smart Contract : }All the interactions among the users and
developers are handled by the smart contract . It check the new uploaded
applications and provide the information about the malware and benign.
Also it store the new information about the new apks. The smart contract
interacts with developer and user to approve the application and notify
the benign or malware.

\begin{lyxlist}{00.00.0000}
\item [{\textbf{Methods:}}] Contracts are structures that define the essence
of the deal. Several contracts have requirements that only require
a certain organization to execute them; others may be accessible to
all participants. The strategies used in the smart contract are directly
related to the effectiveness of the contract.
\item [{\textbf{Modifiers:}}] Modifiers changes the behavior of the application
features. it can only define variables in this block before execution.
it can restrict the access to contract function according to malicious
applications
\item [{\textbf{Variables:}}] Variable holds a value and that value can
change depends on function call or conditions. Based on the smart
contract, variables can be able to store a specific data type.
\end{lyxlist}
\begin{algorithm}
\SetAlgoLined
Contract$ $ is: $ WaitForCheckingMalwaree$\;
Devloper$ $ is: $ ReadtToUploadAP$\;
Approva$ $ is: $ WaitingToSucessOrFail$\;

\eIf{apkHashCheck(distributedLedger)}{
Contract$ $ is: $ sucessSign$\;
Devloper$ $ is: $ sucessProvidedAP$\;
Approve $=$ sucessApproval(If app is not Malware)\;
}
       {
Contract$ $ is: $ denySign$\;
Devloper$ $ is: $ denyProvidedAP$\;
Approve $=$ denyApproval(If app is Malware)\;

    }

\caption{Smart contract approvers uploaded apk}
\end{algorithm}

\begin{algorithm}
\SetAlgoLined
$apk \gets$ DownlodedApp\;
$apkHash \gets$ ApplyHash(apk)\;
Approve$ $ is: $ WaitingToSucessOrFail$\;

\eIf{BlockChainLedger(apkHash)}{
Approve $=$ sucessApproval(Malware information not found)\;
}
       {
Approve $=$ denyApproval\;

    }

\caption{Smart contract approvers download apk}
\end{algorithm}

\section{PERFORMANCE EVALUATION \label{sec:Perf}}

In this section, we discuss the experiment results of our proposed
framework. It include the dataset, evaluation measures, results and
comparison with other works. The proposed model based on deep learning
algorithm and blockchain provides the strong evidence of the results,
which is obtained from the experiments.

\subsection{Dataset}

The dataset that we used contains 18,850 normal Android application
packages and 10,000 malware android packages with different features.
It collected around 13,000 Android application packages (. apk) as
normal apps from different resources and 6971 malicious applications
from known sources such as DroidKin dataset \cite{gonzalez2014droidkin},
Android Malware Genome Project \cite{Zhou2012} and AndroMalShare
\cite{SandDroid}. They extracted the permissions at installation
and run time after running the collected Android application packages
(. apk) using emulator bluestack \cite{Bluestack2020}. In this study,
we used the new version of their dataset that contains 18,850 normal
Android application packages and 10,000 malware.

\subsection{Experimental setup}

In this paper, we extract dynamic and static features. The dynamic
analysis is done in real time devices to check the real time performance
of the network. we utilized 8 mobiles phones with different configurations
, Android 10.0 , 6 GB RAM, Processor Kerin 980, 128 GB ROM. every
smart phone process an average 400 apps daily. All phones contains
sim card with 4G network connection. The execution of the run time
derives are determined when chosen the input generation. Moreover,
to analysis the dynamic behavior of malware activities, use the Emulator
(Android Virtual Device) to extract the dynamic features such as API
calls, Events/Action. After extracting the dynamic and static features,
this paper combine and train the model.

\subsection{Evaluation Measures}

We used Python; the programming language to conduct our experiment.
In order to evaluate malware detection systems efficiency true positive
and false positive rate are used in \cite{Ye2017,Baldi2000} TPR
defined as shown in equation \ref{eq:4} .

\begin{equation}
TPR=\frac{{T_{p}}}{{T_{p}+F_{n}}}\label{eq:4}
\end{equation}

If True Positive (TP) is the sum of correctly recognized malware samples
and False Negative (FN) represents the number of wrongly detected
malware samples that are benign. The recognition rate is also known
as TPR. Eq \ref{eq:5} is defined as false positive rate (FPR). \cite{Ye2017}:

\begin{equation}
FPR=\frac{{F_{p}}}{{F_{p}+T_{n}}}\label{eq:5}
\end{equation}

The malware samples incorrectly identified, and true negative (TN)
is the number of positive samples. FPR is also referred to as the
false alarm rate. Overall Accuracy (ACC): Percentage of correctly
identified applications which is shwon in equation \ref{eq:6}:

\begin{equation}
Accuracy=\frac{{T_{p}+T_{n}}}{{T_{p}+T_{n}+F_{p}+F_{n}}}\label{eq:6}
\end{equation}

\subsection{Results Discussion}

\begin{figure}
\includegraphics[scale=0.4]{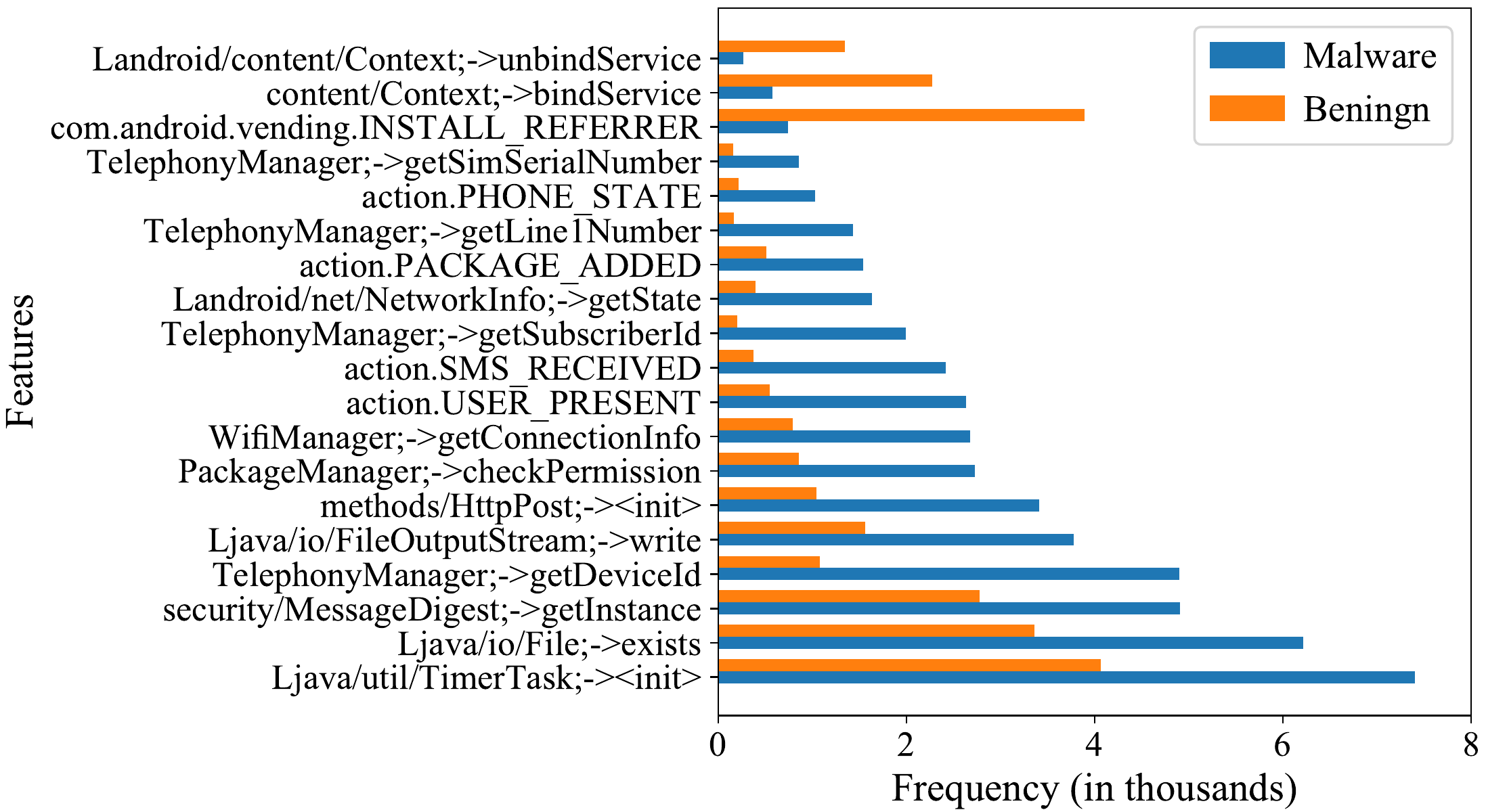}\caption{Top ranked info-gain-based apps use the DroidBot (Permission Excluded)
\label{fig:MB-1}}
\end{figure}

\begin{figure}
\includegraphics[scale=0.4]{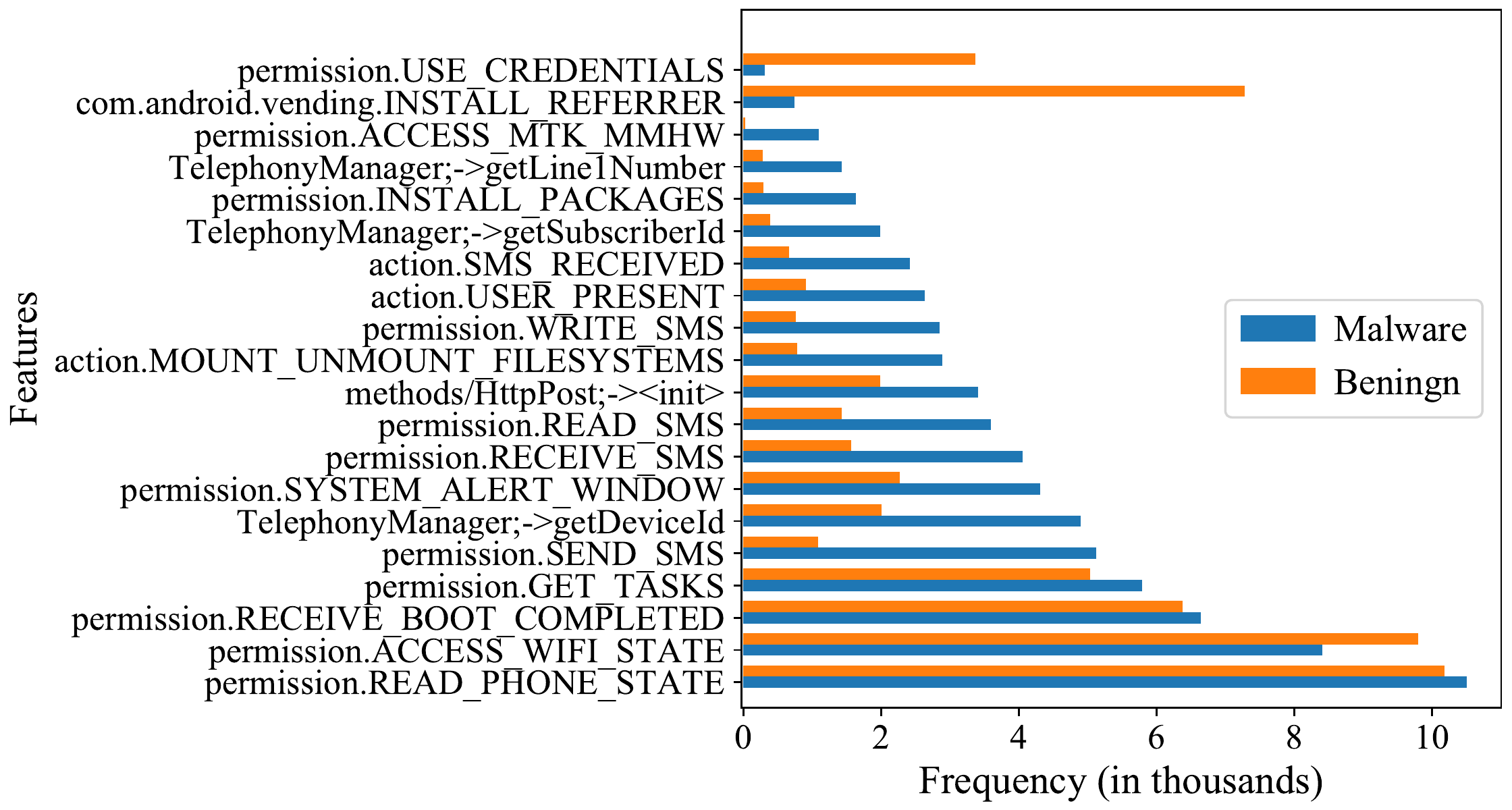}\caption{Top ranked info-gain-based apps use the DroidBot (Permission Included)\label{fig:MB-2}}
\end{figure}

\begin{figure}
\includegraphics[scale=0.36]{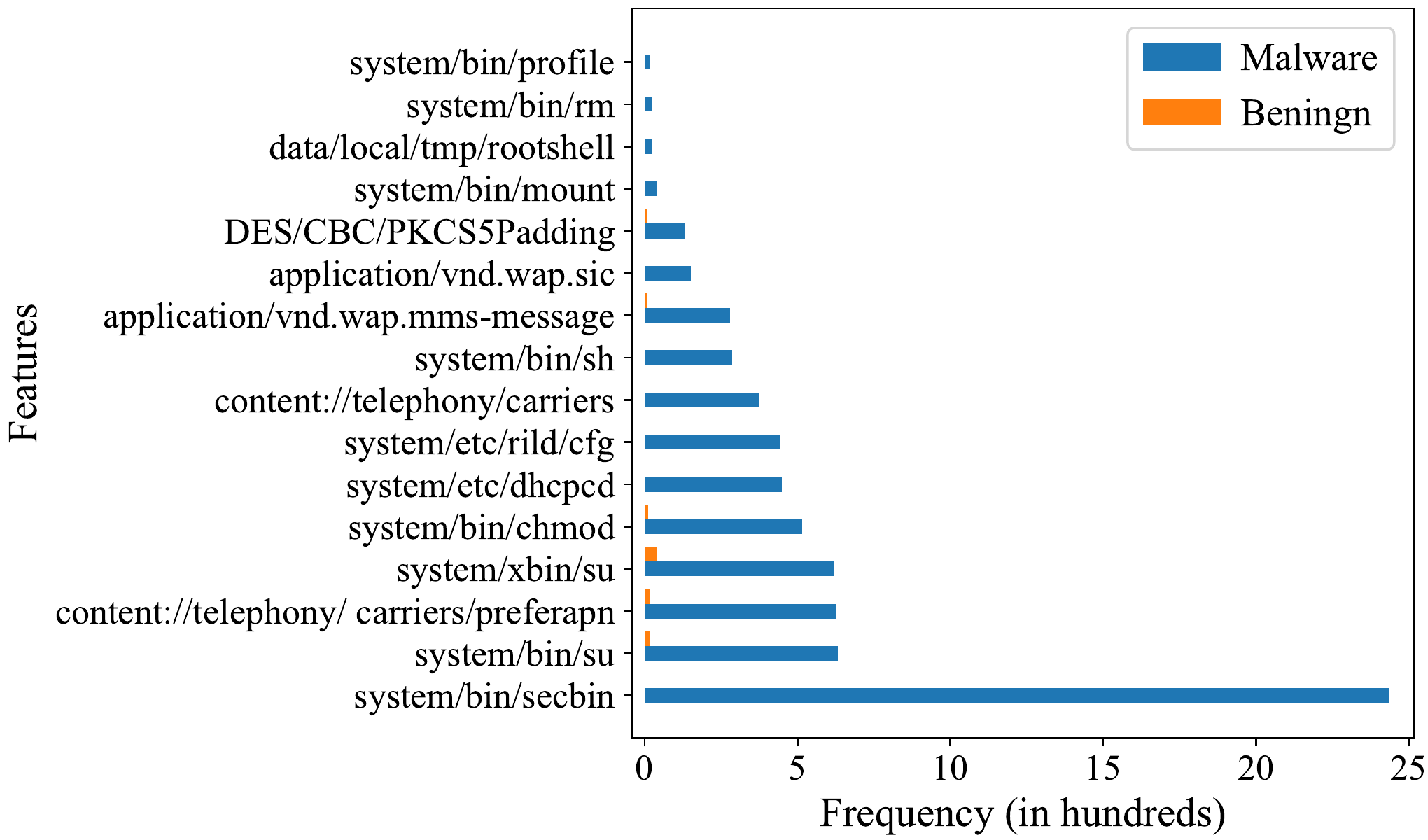}\caption{Top ranked info-gain based Process\label{fig:MB-3}}
\end{figure}

\begin{figure}
\includegraphics[scale=0.42]{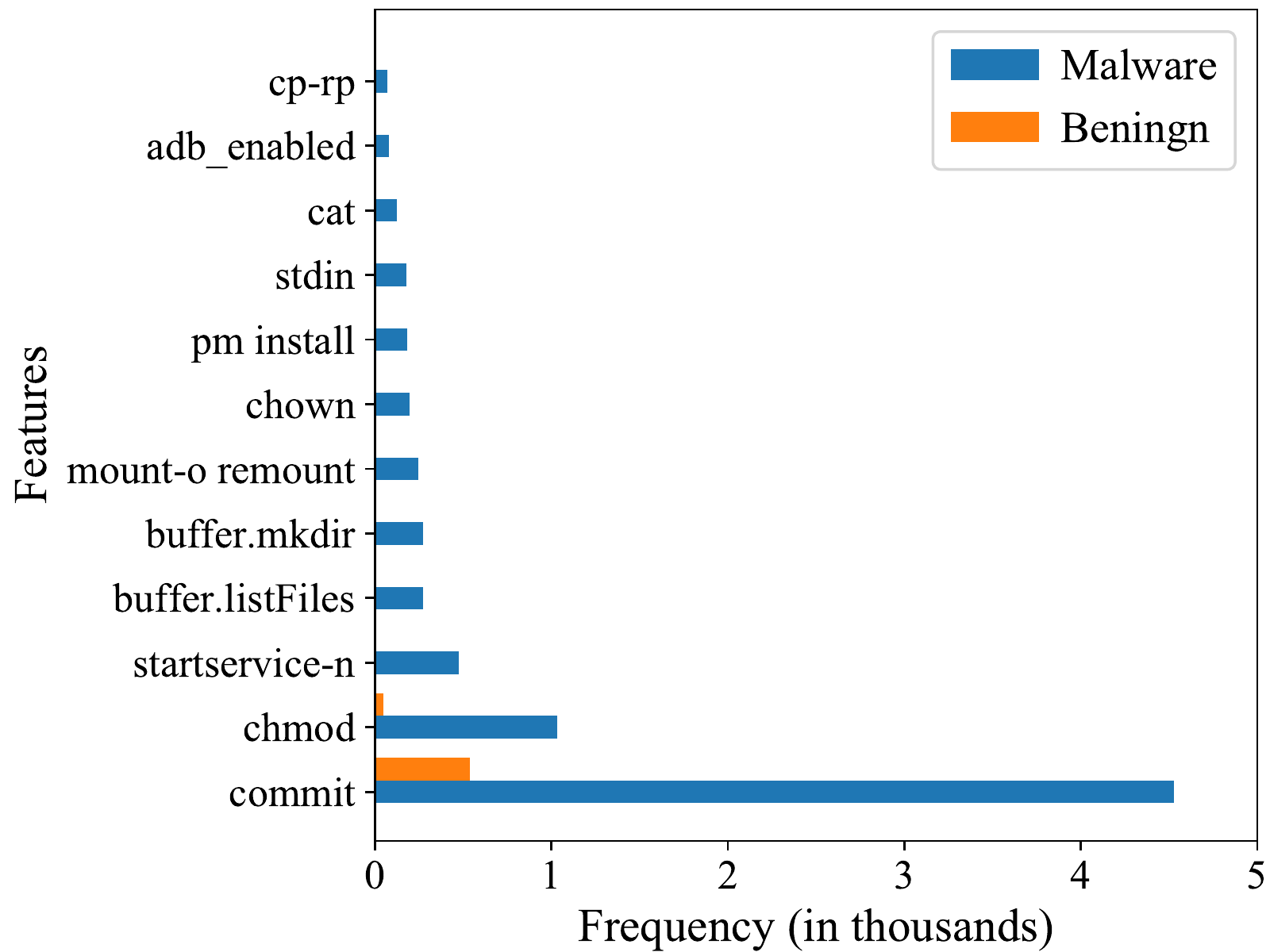}\caption{Top ranked info-gain based Intents\label{fig:MB-4}}
\end{figure}

\begin{figure}
\includegraphics[scale=0.32]{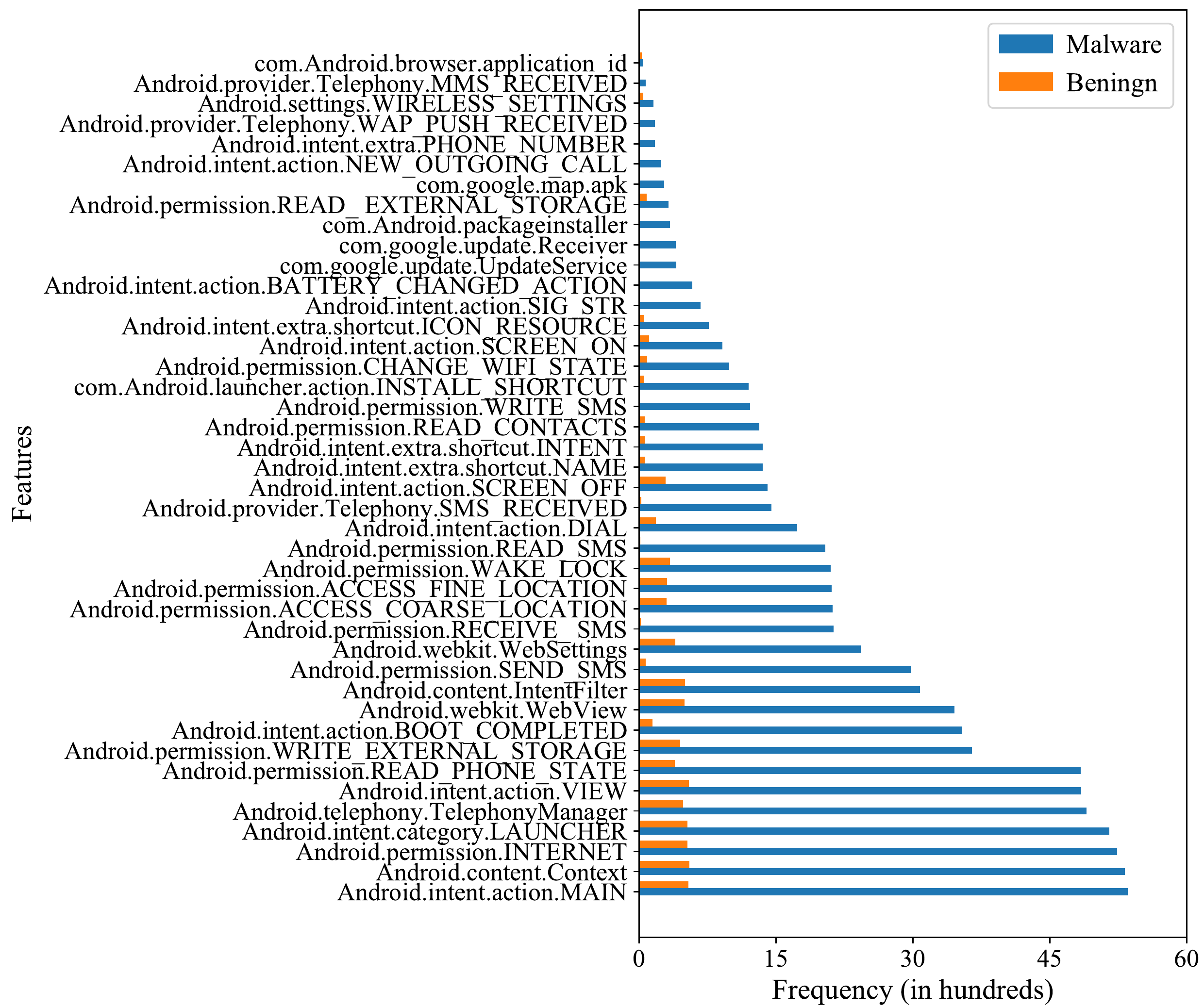}\caption{Top ranked info-gain based Intents\label{fig:MB-5}}
\end{figure}

The detailed empirical study of our proposed research work contains
two observations. Initially, our key goal is to build a deep learning
based model which enable to identify and diagnose Android malware
and benign applications. 
For this purpose, we analyze static and dynamic
features based on overall information gain score in 
 At first, inspected information gain based on
frequency count of features such as permissions, connections, intents
of the static and dynamic features are exploited, as depicted in Figure
\ref{fig:MB-1}-\ref{fig:MB-5}. From the observed behavior of mentioned
features, we can conclude that API calls and services frequency ratio
have inverse relationship between benign and malware applications.
Whereas specifically, as shown in Figure \ref{fig:MB-5}, frequnecy
count of Intents is much less in benign compared with malware applications.
In contrast, the utilization of permission does not help to distiguish
between benign and malware due to same frequency count, as shown in
Figure \ref{fig:MB-2}. Therefore, we exploit useful static and dynamic
features to construct a high quality model for training.






\begin{figure}
	\centering
\includegraphics[scale=0.35]{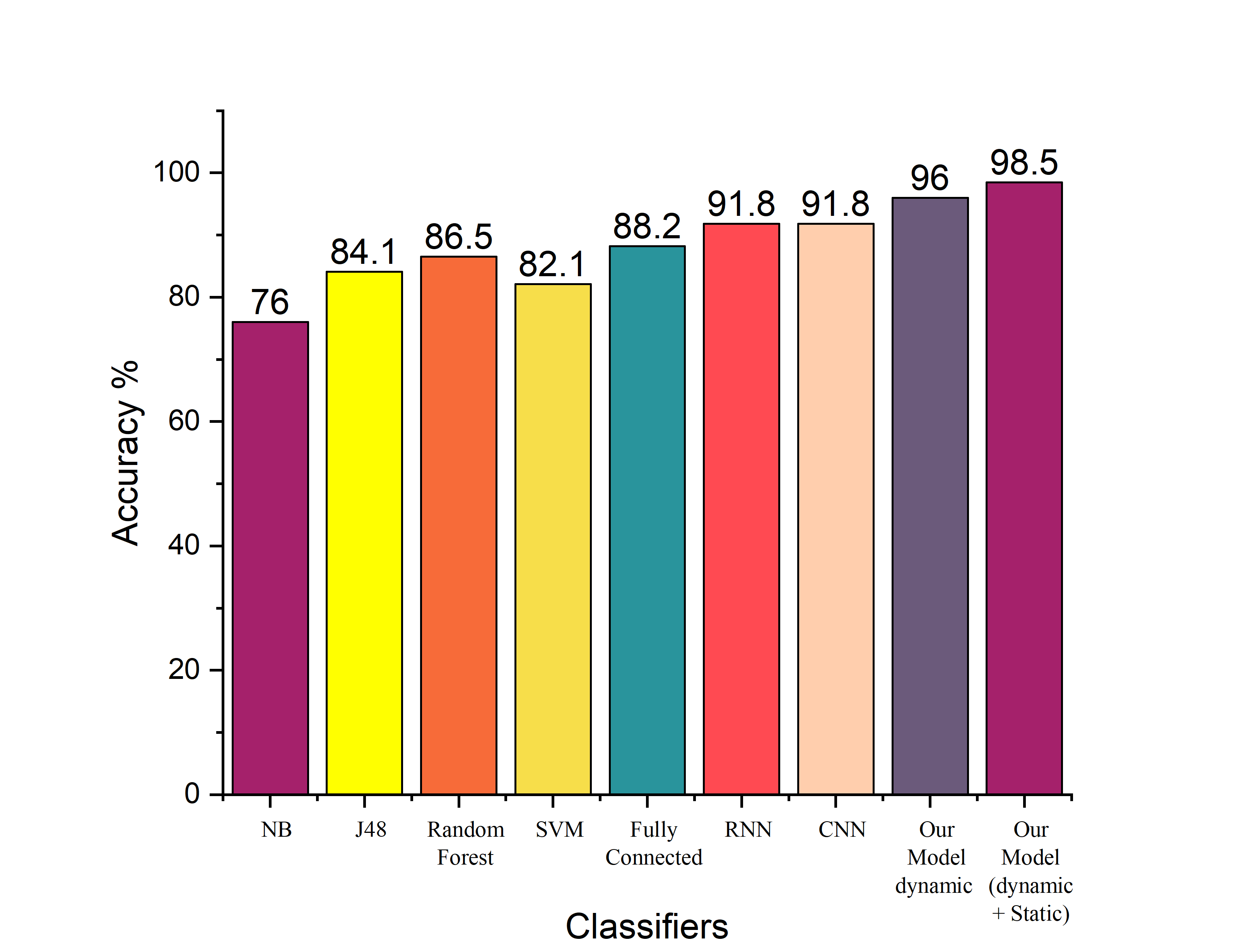}\caption{Comparison between machine and deep learning classifiers\label{fig:Accuracy}}
\end{figure}

\begin{figure}
\includegraphics[scale=0.35]{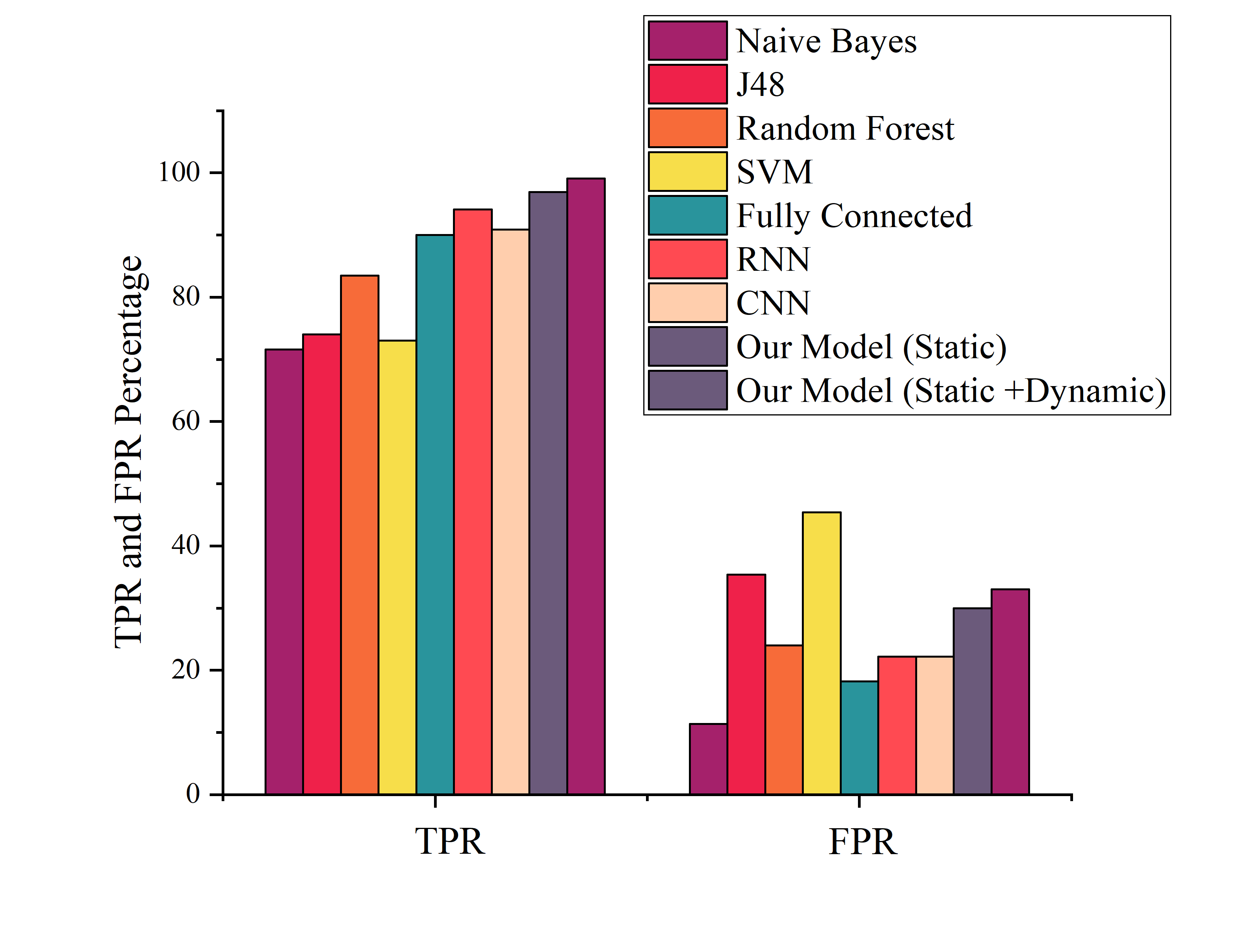}\caption{True positive and false positive performance between different classifiers\label{fig:TPR-FPR}}
\end{figure}

The engineered features are then used to train different algorithms
including Support Vector Machine (SVM), J48, Naive Bayes (NB), Random
Forest, Recurrent Neural Network (RNN), Convolutional Neural Network
(CNN), Fully Connected Deep Neural Network (FC-DN) and compared with
our proposed deep learning model. For ground level evaluation, we
report TPR and FPR of all algorithms \ref{fig:TPR-FPR}. As shown
in Figure \ref{fig:Accuracy}, the proposed approach outperfomed the
previous algorithms by gaining high TPR and accuracy. However, due
to confilcting relationship of features between benign and malware,
the reported FPR is not better than other approaches except SVM and
J48.

Table \ref{tab:hiddenLayers} shows the performance of the deep learning
model with a different combination of hidden layers. These presented
results show only dynamic features using the emulator. We apply different
layers of neurons to compare the best performance of the deep learning
model. Table \ref{tab:hiddenLayers} applied two, three and four layers
of deep learning model, the combination of 200, 200, 200 neurons achieved
best compare with other layers and neurons. Similarly, we repeated
the same experiment to combine the static and dynamic features shown
in Table \ref{tab:HiddenLayers-static-dynamic}. However, this also
has the same layers and neurons. The combination of 200,200,200 in
3 layers achieved better performance than other layers and neurons.

\begin{table}
\caption{Proposed deep learning model with different hidden layers for dynamic
features only\label{tab:hiddenLayers}}

\begin{tabular}{|>{\centering}p{0.05\columnwidth}|c|c|c|c|>{\centering}p{0.1\columnwidth}|}
\hline
No. of layers & No. of Neurons & TPR & FPR & Accuracy & Running time (min:sec)\tabularnewline
\hline
\hline
2 & 200,200 & 0.9663 & 0.337 & 0.9449 & 06:31\tabularnewline
\hline
2 & 400,400 & 0.9903 & 0.2062 & 0.895044 & 13:44\tabularnewline
\hline
\textbf{3} & \textbf{200,200,200} & \textbf{0.9719} & \textbf{0.0728} & \textbf{0.9624} & \textbf{09:05}\tabularnewline
\hline
3 & 400,400,400 & 0.7219 & 0.0158 & 0.8183 & 20:40\tabularnewline
\hline
4 & 200,200,200,200 & 0.9744 & 0.0891 & 0.9508 & 10:39\tabularnewline
\hline
4 & 400,400,400,400 & 0.9622 & 0.115 & 0.9339 & 29:28\tabularnewline
\hline
\end{tabular}
\end{table}

\begin{table}
\caption{Proposed deep learning model with different hidden layers for Static
and dynamic features{\label{tab:HiddenLayers-static-dynamic}}}

\begin{tabular}{|>{\centering}p{0.05\columnwidth}|c|c|c|c|>{\centering}p{0.1\columnwidth}|}
\hline
No. of layers & No. of Neurons & TPR & FPR & Accuracy & Running time (min:sec)\tabularnewline
\hline
\hline
2 & 200,200 & 0.9661 & 0.1229 & 0.9332 & 14:51\tabularnewline
\hline
2 & 400,400 & 0.972 & 0.0918 & 0.9484 & 30:50\tabularnewline
\hline
\textbf{3} & \textbf{200,200,200} & \textbf{0.9956} & \textbf{0.033} & \textbf{0.985} & \textbf{17:21}\tabularnewline
\hline
3 & 400,400,400 & 0.9764 & 0.0834 & 0.9543 & 38:16\tabularnewline
\hline
4 & 200,200,200,200 & 0.9757 & 0.0793 & 0.955 & 20:05\tabularnewline
\hline
4 & 400,400,400,400 & 0.9717 & 0.0907 & 0.9486 & 43:52\tabularnewline
\hline
\end{tabular}
\end{table}

Table \ref{tab:Time-deep-models} compare the time for different deep
learning models. Experiment results indicates that proposed model
reduce the computational time and also achieve detection performance
for the Android IoT devices.

\begin{table}
\caption{Time comparison of deep learning model construction \label{tab:Time-deep-models} }

\begin{tabular}{|c|c|c|c|}
\hline
No. of layers & No. of Neurons & TPR & Time \tabularnewline
\hline
\hline
3 & 200,200,200 & Fully Connected & 140\tabularnewline
\hline
3 & 200,200,200 & RNN & 135\tabularnewline
\hline
3 & 200,200,200 & CNN & 120\tabularnewline
\hline
3 & 200,200,200 & Our Proposed (Static) & 96\tabularnewline
\hline
3 & 200,200,200 & Our Proposed (Dynamic) & 99\tabularnewline
\hline
\end{tabular}
\end{table}


\begin{table}
		\centering
	\caption{PERFORMANCE OF FEDERATED AND BLOCKCHAIN-FEDERATED METHODS}
	\label{tab:PERFORMANCE-OF-CENTRALIZED-AND-FEDERATED-METHODS}
	\begin{tabular}{|c|c|c|c|c|}
		\hline 
		Train on & TPR & FPR & Accuracy & Time in seconds\tabularnewline
		\hline 
		\hline 
		Local user 1 & 97.41 & 17.92 & 93.95 & \multirow{5}{*}{189.05}\tabularnewline
		\cline{1-4} \cline{2-4} \cline{3-4} \cline{4-4} 
		Local user 2 & 96.15 & 13.71 & 93.59 & \tabularnewline
		\cline{1-4} \cline{2-4} \cline{3-4} \cline{4-4} 
		Local user 3 & 95.70 & 13.13 & 93.34 & \tabularnewline
		\cline{1-4} \cline{2-4} \cline{3-4} \cline{4-4} 
		Local user 4 & 95.66 & 14.22 & 93.41 & \tabularnewline
		\cline{1-4} \cline{2-4} \cline{3-4} \cline{4-4} 
		Local user 5 & 96.47 & 14.70 & 93.64 & \tabularnewline
		\hline 
		Federated & 97.34 & 12.41 & 95.35 & 0.052\tabularnewline
		\hline 
		Blockchain-Federated & 98.64 & 13.21 & 98.05 & times\tabularnewline
		\hline 
	\end{tabular}
\end{table}

In Table \ref{tab:PERFORMANCE-OF-CENTRALIZED-AND-FEDERATED-METHODS}, we conduct the classification of Android Malware based
on federated learning and federated learning model for blockchain
with 5 users. The training of local model parameters of local epochs
is 30 and batch size is 1. We select all kinds of features such as
API calls, permissions and  intents. The performance of local users
and federated learning model shown in Table . The federated learning
takes less time then train the local model. The client required 0.93
mili seconds to send model from client to server. Thus the proposed
framework has less communication time. In the way blockchain and deep
learning model aggregate the different features i.e., permission,
intent, dynamic features and static feature and perform well then
other previous approaches.

\begin{table}
	\centering
	\caption{PERFORMANCE OF DIFFERENT NUMBER OF CLIENTS}
	\label{tab:PERFORMANCE-OF-DIFFERENT-NUMBER-OF-CLIENTS}
	\begin{tabular}{|c|c|c|c|}
		\hline 
		Train on & TPR & FPR & Accuracy\tabularnewline
		\hline 
		\hline 
		2 clients & 96.45 & 13.67 & 96.99\tabularnewline
		\hline 
		5 clients & 97.54 & 12.49 & 97.45\tabularnewline
		\hline 
		10 clients & 97.43 & 12.18 & 97.74\tabularnewline
		\hline 
		15 clients & 97.72 & 13.78 & 97.98\tabularnewline
		\hline 
		30 clients & 97.43 & 12.41 & 97.98\tabularnewline
		\hline 
		40 clients & 96.54 & 11.89 & 97.45\tabularnewline
		\hline 
	\end{tabular}
\end{table}

In Android platform there are many clients use many kind of services,
so we test the classification performance based on number of clients.
Each user choose different 10,000 features sets randomly form the
training dataset to train the local model with various features. According
to Table \ref{tab:PERFORMANCE-OF-DIFFERENT-NUMBER-OF-CLIENTS} the number of clients are increases then the accuracy will
increase. Therefore, more users, the model can perform better.

\begin{table}
	\centering
	\caption{PERFORMANCE OF DIFFERENT TYPES OF FEATURES}
	\label{tab:PERFORMANCE-OF-DIFFERENT-TYPES-OF-FEATURES}
	\begin{tabular}{|c|c|c|c|}
		\hline 
		Feature & TPR & FPR & Accuracy\tabularnewline
		\hline 
		\hline 
		API calls & 95.12 & 20.34 & 91.41\tabularnewline
		\hline 
		Permission & 85.90 & 31.14 & 81.43\tabularnewline
		\hline 
		Intents & 92.34 & 19.24 & 92.12\tabularnewline
		\hline 
		Federated & 98.64 & 20.75 & 94.42\tabularnewline
		\hline 
		Federated Blockchain & 99.10 & 17.87 & 98.62\tabularnewline
		\hline 
	\end{tabular}
\end{table}

In Android malware datasets have many different features to detect
the malware, therefore we classify the different features set in Table
\ref{tab:PERFORMANCE-OF-DIFFERENT-TYPES-OF-FEATURES} and compare with the federated learning.

Furthermore, we focus on blockchain integration functionality among
the smart contract, hyper ledger, deep learning and users. we implemented
the Ethereum smart contract using Remix IDE. All roles have been tested
to ensure that the smart contract's worked properly. The developer
uploads the apk files to the blockchain network and stores the hash
in to the smart contract. In Remix, for different role different address
are stores such as user (0 \texttimes ca35b7d915458ef540ade6d35458dfe2f44e8fa733c)
and developer (0 \texttimes 18965a09acff6d2a60 dcdf8bb4aff308fddc180c),
to test the smart contract code. Functions are designed to approve
or deny the Android apps. 

\begin{figure}
\includegraphics[scale=0.35]{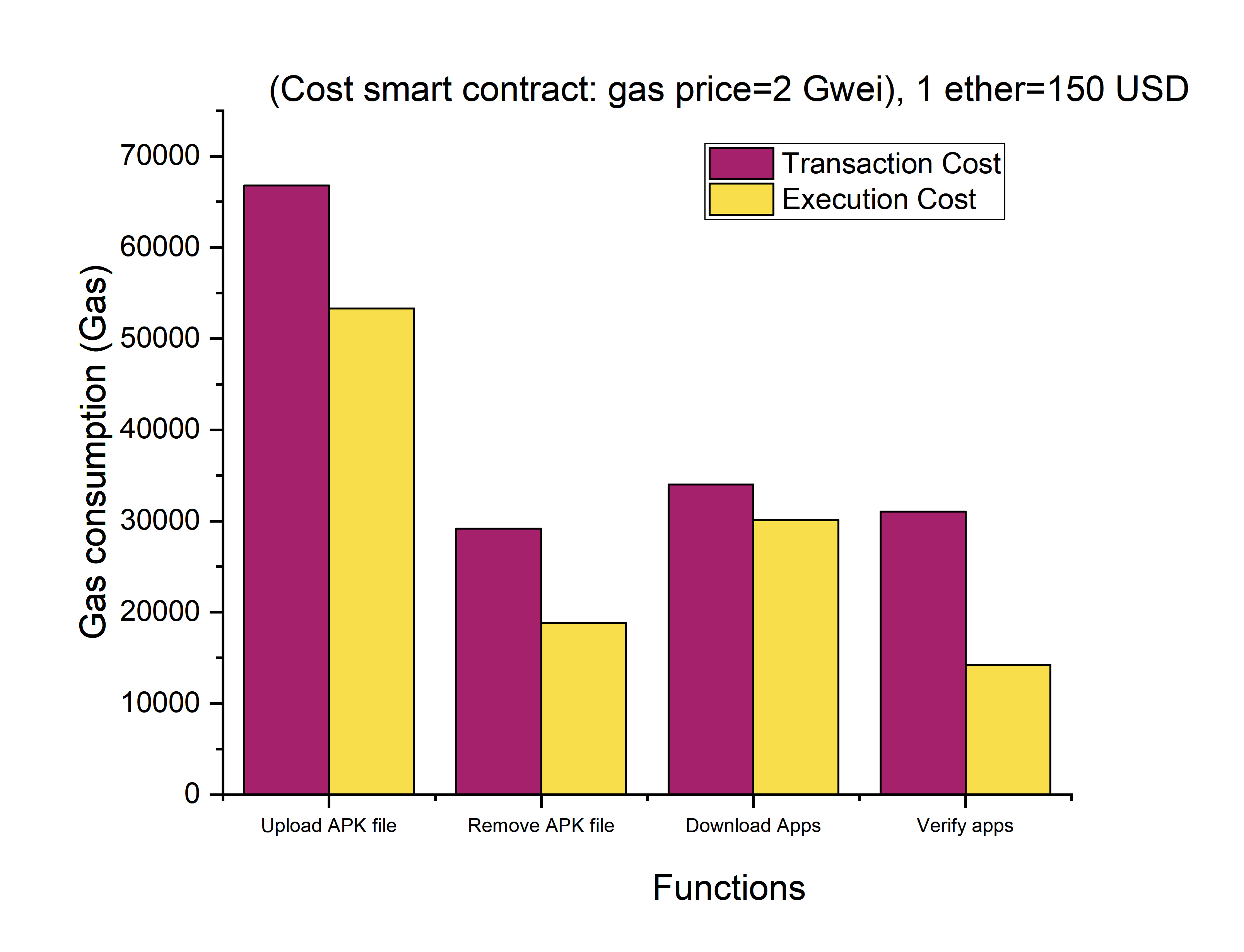}\caption{Gas consumption for downloading and uploading file\label{fig:GAS-Execuation-cOST}}
\end{figure}
\begin{figure}
\includegraphics[scale=0.35]{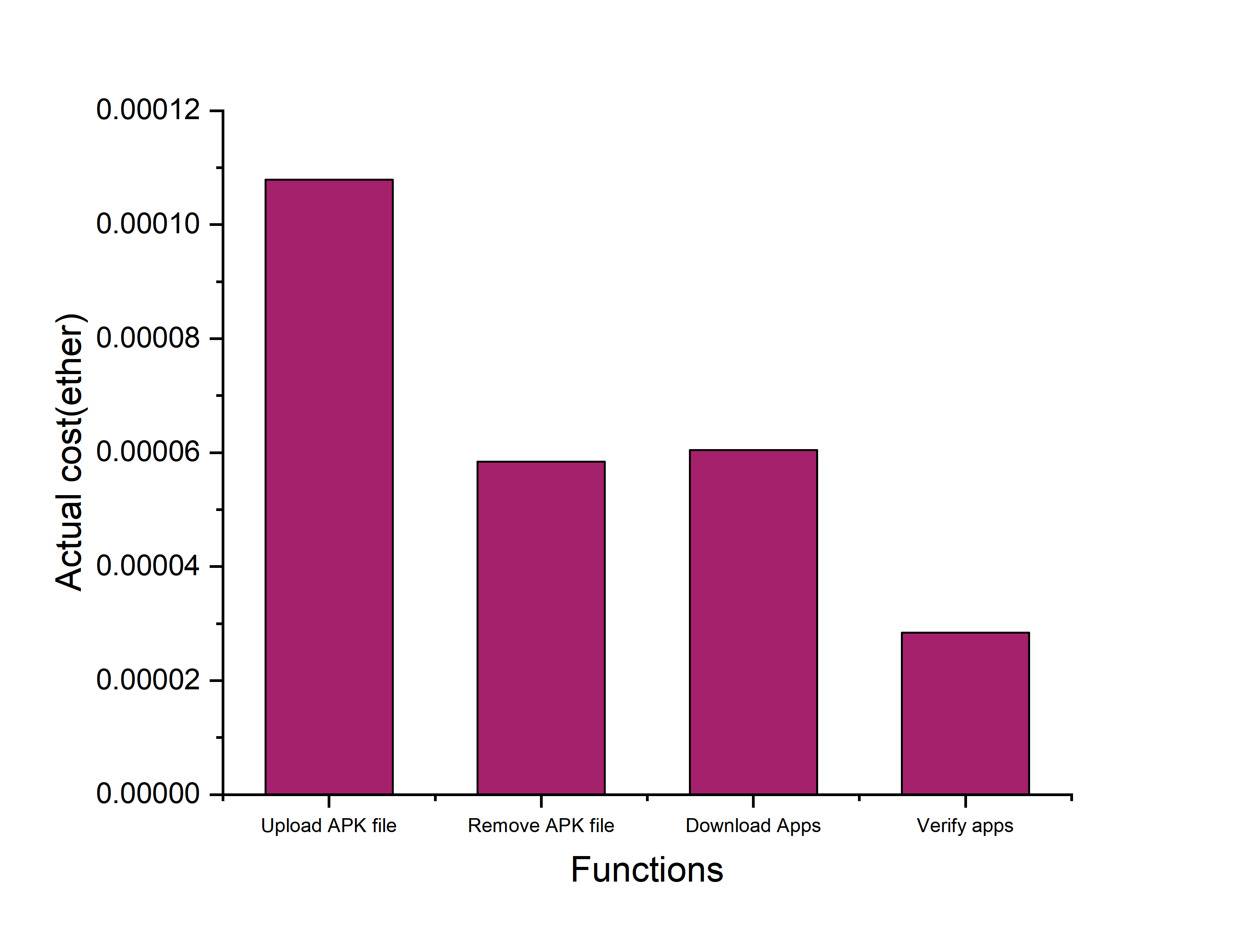}\caption{Actual cost for downloading and uploading file\label{fig:GAS_COST}}
\end{figure}

To utilize the IPFS storage, the smart contract transaction and gas
execution are recorded as 1808246 and 1338218 respectively. The transaction
cost is required to upload the Android apps and the amount of gas
is necessary to verify the hash values of the harmful apps. When uploading
the apps in the server, the execution cost of the function is USD
0.016. When an Android app is downloading USD 0.0088 is required.
When verify the harmful apps the price measured USD 0.0091. Whenever
the virus found to form the file then the hash values are stored in
the blockchain. Furthermore, the execution and Either cost measured
in Figure \ref{fig:GAS-Execuation-cOST} and \ref{fig:GAS_COST} respectively.

\begin{figure*}[t]
	\centering
\includegraphics[scale=0.25]{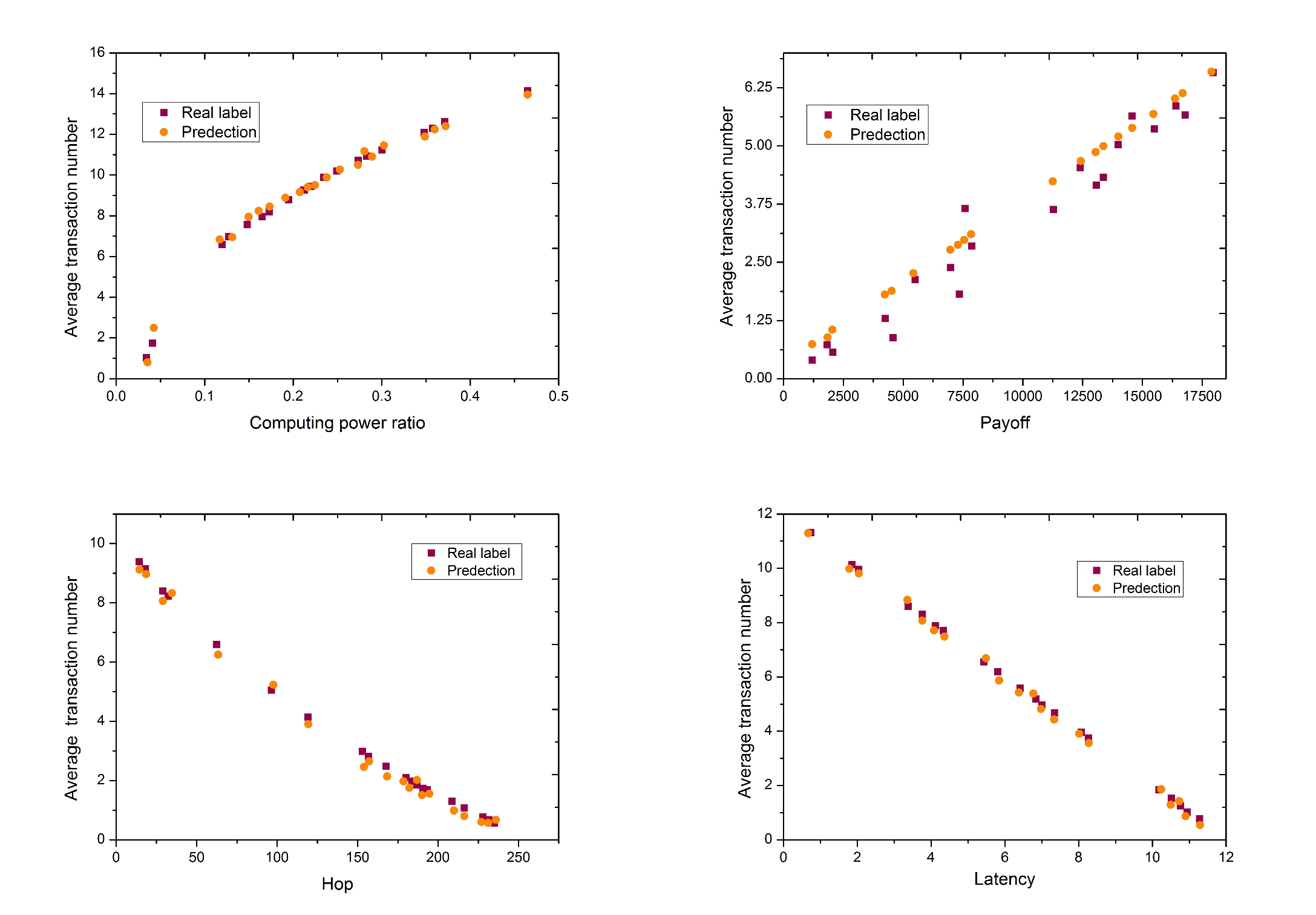}\caption{Correlation between the computing power ratio and average transaction\label{fig:Correlation-among-the}}
\end{figure*}

The Figure \ref{fig:Correlation-among-the} shows the simulation results
to prove the combination of blockchain and deep neural network increases
the performance in terms of reducing the computational cost of the
neural network. It shows the training labels increase the prediction
performance of the deep learning model is increase. Moreover, the
blockchain network calculates the value of nodes by using the deep
neural network. And then selects the nodes and calculate the threshold
value the features of the dataset taken as the input information to
the blockchain nodes. The mining pools provides output to the users.
The average number of transaction shown in the Figure \ref{fig:Correlation-among-the}.
The blue label defines the real labels, and the red provides the prediction.
Figure \ref{fig:Correlation-among-the} (a) the correlation among
the computing power ratio and the average transaction is defined as
the trend of the blue dots, and red dots show an increasing pattern
of the average transaction with the computing power ratio, which is
constant with the real world decentralized network. Figure \ref{fig:Correlation-among-the}
(b) demonstrates the payoff increases, more transactions the node
will have. Figure \ref{fig:Correlation-among-the} (c) and (d) is
the nodes are negatively correlated

\subsection{Comparison with other works}

In this section, we evaluate the efficiency of the proposed deep learning
model, and we compare state-of-art deep learning and machine learning
approaches. As we can see in Figure \ref{fig:Accuracy} compares the
accuracy with machine learning and deep learning classifiers. Moreover,
we also compare our deep learning model with previous literature shown
in Table \ref{tab:Compersion-accuracy-with}. Table \ref{tab:Compersion-accuracy-with}
hows that our method achieve higher accuracy than other techniques.

Furthermore, we compare our work with the \cite{feng2020performance} and \cite{Kumar2019}
method which introduces the blockchain with Android malware detection.
It only proposes and stores the information of malware that is not
able to the real-time deployment of blockchain. On the other hand,
compared to \cite{feng2020performance} and \cite{Kumar2019} our solution has
better achievement to secure the IoT devices. Additionally, the Table
\ref{tab:compare-blockchain} shows the results of the comparative
analyses of the blockchain applications. As can be seen in our contribution,
a blockchain application is used to identify whether benign or malware
when uploading and downloading the apps from the Internet.

\begin{table}
\begin{tabular}{|c|c|>{\centering}p{0.1\columnwidth}|c|c|}
\hline
Authors & Algorithm & Capacity for feature diversity & Accuracy & F-measure\tabularnewline
\hline
\hline
OURs & Proposed  & High & 96\% & 0.98\tabularnewline
\hline
\cite{Kim2019} & DNN/RNN & medium & 90\% & NA\tabularnewline
\hline
\cite{McLaughlin2017a} & CNN & low & 90\% & NA\tabularnewline
\hline
\cite{zhu2020sedmdroid} & Multi-Layer Perception & low & 89\% & 0.89\tabularnewline
\hline
\cite{taheri2020similarity} & KMNN/ ANN/ FNN  & High & 90 \%  & NA\tabularnewline
\hline
\cite{feng2020performance} &  RNN and LSTM  & low & 96\% & NA\tabularnewline
\hline
\cite{zhang2019familial} & DNN & High & 93.9 & NA\tabularnewline
\hline
\cite{chen2015finding} & Bayesian & low & 92\% & NA\tabularnewline
\hline
\cite{Yuan2016a} & SVM & low & NA & 0.98\tabularnewline
\hline
\cite{Huang2013} & Graph Based &  NA & 95.4\% & NA\tabularnewline
\hline
\end{tabular}\caption{performance comparisin with other state of the art approaches \label{tab:Compersion-accuracy-with}}
\end{table}

\begin{table}
\begin{tabular}{|>{\raggedright}p{0.15\columnwidth}|>{\raggedright}p{0.15\columnwidth}|>{\raggedright}p{0.15\columnwidth}|>{\raggedright}p{0.15\columnwidth}|>{\raggedright}p{0.15\columnwidth}|}
\hline
\textbf{Primitive} & \textbf{Block Verify \cite{BlockVerify2019}} & \textbf{Sigma Ledger \cite{Alzahrani2018}} & \textbf{Stop TheFake \cite{StopTheFakes2018}} & \textbf{This Work}\tabularnewline
\hline
\hline
Blockchain & Private & Private & Private & Private\tabularnewline
\hline
Target & Goods & Goods & Picture, Video & Android APK\tabularnewline
\hline
Function & Detect, Identify & Tag, Detect & Detect, Record & Detect, Identify\tabularnewline
\hline
Smart Contract & Product Label & QRCode, RFID & Copyright, Catalog & Hash, Feature of APK\tabularnewline
\hline
\end{tabular}\caption{Compare with other blockchain techniques\label{tab:compare-blockchain}}
\end{table}

\section{Conclusion \label{sec:Con}}

In this paper, a new approach is presented to integrate the blockchain
and multi-level deep learning model for the detection of malware activities
in a real-time environment, especially for Android IoT devices. Our
proposed framework works as follows: 1) Dveloper creates a malware
2) Multi-level deep learning model distributes the malware features
into various cluster and chooses best deep learning model for each
cluster. 3) Moreover, it makes decisions by analyzing the previous
data which is already stored in blockchain distributed ledger and
stores the new features of the malware activities in the blockchain
4) Finally, the blockchain smart contract provides the notification
(of malware) to the user regarding verification of Android app during
uploading or download process. To achieve better security for IoT
devices regarding malware detection in realtime environments, millions
of Android application features (malware and benign) were stored in
the blockchain database. Therefore, we designed a multi-layer deep
learning model for a large number of malware and benign features that
incept the malicious application for Android IoT devices. The proposed
model supports the multiple levels of clustering for single data distribution.
Furthermore, the smart contract verifies the malicious application
to uploading and downloading the Android apps through the network.
It can approve or deny to uploading and downloading harmful Android
applications. The proposed model can identify the malware effectively
which can provide more security for the Android IoT devices.

\bibliographystyle{ieeetr}
\bibliography{j9}

\end{document}